\documentclass[%
 aip,
 amsmath,amssymb,
 reprint,
]{revtex4-1}

\usepackage{graphicx}% Include figure files
\usepackage{dcolumn}% Align table columns on decimal point
\usepackage{bm}% bold math
\usepackage[utf8]{inputenc}
\usepackage[T1]{fontenc}
\usepackage{mathptmx}
\usepackage{etoolbox}
\usepackage{url}
\usepackage{xcolor,comment}
\usepackage{booktabs}
\usepackage[caption=false]{subfig}
\graphicspath{ {./} }

\makeatletter
\def\@email#1#2{%
 \endgroup
 \patchcmd{\titleblock@produce}
  {\frontmatter@RRAPformat}
  {\frontmatter@RRAPformat{\produce@RRAP{*#1\href{mailto:#2}{#2}}}\frontmatter@RRAPformat}
  {}{}
}%
\makeatother
\begin{document}

\preprint{AIP/123-QED}

\title{Decomposing the Dynamics of the Lorenz 1963 model using Unstable Periodic Orbits: Averages, Transitions, and Quasi-Invariant Sets}
% Force line breaks with \\
\author{Chiara Cecilia Maiocchi}
 \altaffiliation{Corresponding author. Email: c.maiocchi@pgr.reading.ac.uk}
\author{Valerio Lucarini}
\altaffiliation{Email: v.lucarini@reading.ac.uk}
\affiliation{Centre for the Mathematics of Planet Earth, University of Reading, RG6 6AH, United Kingdom}
\affiliation{Department of Mathematics and Statistics, University of Reading, RG6 6AH, United Kingdom}
\author{Andrey Gritsun}
\altaffiliation{Email: asgrit@mail.ru}
\affiliation{Institute of Numerical Mathematics, Russian Academy of Sciences, Moscow, 119333, Russia}

\date{\today}% It is always \today, today,

\begin{abstract}
Unstable periodic orbits (UPOs) are a valuable tool for  studying chaotic dynamical systems, as they allow one to distill their dynamical structure. We consider here the Lorenz 1963 model with the classic parameters' value. We investigate how a chaotic \textcolor{black}{trajectory} can be approximated using a complete set of UPOs up to  symbolic dynamics' period 14. At each instant, we rank the UPOs according to their proximity to the position of the orbit in the phase space. We  study this process from two different perspectives. First, we find that longer period UPOs overwhelmingly provide the best local approximation to the trajectory. Second, we construct a finite-state Markov chain by studying the scattering of the orbit between the neighbourhood of the various UPOs. Each UPO and its neighbourhood are taken as a possible state of the system. Through the analysis of the subdominant eigenvectors of the corresponding stochastic matrix we provide a different interpretation of the mixing processes occurring in the system by taking advantage of the concept of quasi-invariant sets. 
\end{abstract}

\maketitle

\begin{quotation}
The attractor of a chaotic system is densely populated by an infinite number of unstable periodic orbits (UPOs), which are exact periodic solutions of the evolution equations. UPOs can be used to decompose the complex phenomenology of a chaotic flow into elementary components and have shown great potential for the understanding of macroscopic features in turbulent fluid flows. Here we investigate how a long forward trajectory of the celebrated Lorenz 1963 model featuring the classical parameters' value can be seen as a scattering process where the scatterers are the UPOs. This process helps \textcolor{black}{elucidate} how a generic ensemble of initial conditions converges to the invariant measure through diffusion and provide a new interpretation of quasi-invariant sets of the system in terms of UPOs.  
\end{quotation}

\section{Introduction}
%\subsection{UPOs are the building blocks of chaotic dynamics}
\label{buildingblocks}
\textcolor{black}{Unstable periodic orbits (UPOs) play an important role in the analysis of dynamical systems that exhibit chaotic behaviour. \textcolor{black}{As noticed early on by Poincar\'e, \cite{poincare_1893} UPOs provide a powerful framework for understanding their statistical properties\cite{cvitanovic_1991}} (see ChaosBook \cite{cvitanovic_2005} for an extensive discussion of this.)}
UPOs can be considered as islets of order in a landscape of chaos and can be used to reconstruct the statistical properties of a chaotic dynamical system \cite{grebogi_1988}. In fact, \textcolor{black}{when} UPOs are dense in the attractor \cite{eckmann_1985}, they can approximate with an arbitrary accuracy any trajectory in the system on the attractor  \cite{bowen_1975}. \textcolor{black}{This is because}  the trajectory is continuously repelled from the neighbourhood of one UPO to another, as a result of the instability of the UPOs. Within this context it is possible to develop a theory that allows dynamical averages to be written as weighted sums over the full set of UPOs. \textcolor{black}{Gutzwiller} \cite{gutzwiller_2013} first demonstrated  that UPOs are the essential building blocks of chaotic dynamics. Cvitanovi\'c \cite{cvitanovic_1988} argued that UPOs are the optimal practical tool for measuring the invariant properties of a dynamical system.  Ruelle later derived the dynamical $\zeta$ function \cite{ruelle_2004} , \textcolor{black}{that allows one to write averages over the invariant measure of the system as a weighted sum over the infinite set of UPOs}. 

These results are proven to be valid for dynamical systems exhibiting strong chaoticity \cite{ruelle_1999,katok_1997}, such as uniformly hyperbolic and Axiom A  systems \cite{smale_1967, bowen_1972}. 
However, in complex models of fluid flows,  it is often difficult, if not impossible, to verify the hypothesis required for the validity of periodic orbit expansion. 
When turbulent conditions are considered, such systems live, after transients have died out, in nonequilibrium steady state (NESS) \cite{gallavotti_2014}. This state is in general characterised by generation of entropy, contraction of phase space and finite-time predictability.
The \textit{'Chaotic hypothesis'} of Gallavotti offers a possible solution to the first problem, allowing to consider \textit{'a turbulent fluid as a transitive Axiom A system for the purpose of computing macroscopic properties of the system'} \cite{gallavotti_1998,gallavotti_1995}.

It is usually assumed that considering short period UPOs allows for a sufficiently accurate estimate of ergodic averages \cite{eckhardt_1994,cvitanovic_1995,artuso_1990}. \textcolor{black}{Indeed, some authors have attempted to define what is the optimal choice of low-period UPOs for approximating ergodic averages of given observables for both discrete \cite{Hunt1996} and continuous-time \cite{Yang2000} dynamical systems. Note that, instead, Zoldi and Greenside \cite{Zoldi1998} have emphasized that in some cases long-period UPOs are essential for achieving good accuracy when performing averages. Along these lines, Lasagna \cite{lasagna_2018, lasagna_2020} found numerical evidence that long period UPOs could be used as accurate proxies of chaotic trajectories. His proposal, in contrast with the previous authors, is that few long UPOs might be able to capture the statistical properties of chaotic trajectories. %, so that one should focus the computational resources to seeking few high-period UPOs.
%In line with this work, we find that longer UPOs have the lion's  share in reproducing the invariant measure of the system. 
One should keep in mind that the efficient computation of UPOs in high dimensional systems still represents an open challenge \cite{chandler_2013}.}

\begin{comment}These results are proven to be valid for dynamical systems exhibiting strong chaoticity \cite{ruelle_1999,katok_1997}, such as hyperbolic and Axiom A  systems \cite{smale_1967, bowen_1972}. 
However, in complex models of fluid flows,  it is often difficult, if not impossible, to verify the hypothesis required for the validity of periodic orbit expansion. 
When turbulent conditions are considered, such systems live, after transients have died out, in nonequilibrium steady state (NESS) \cite{gallavotti_2014, lucarini_2014}. This state is in general characterised by generation of entropy, contraction of phase space and finite-time predictability.
The \textit{'Chaotic hypothesis'} of Gallavotti offers a possible solution to the first problem, allowing to consider \textit{'a turbulent fluid as a transitive Axiom A system for the purpose of computing macroscopic properties of the system'} \cite{gallavotti_1998,gallavotti_1995}. \\
\end{comment}

\subsection{Unstable Periodic Orbits: Applications}
A first application of periodic orbit expansion was performed by Auerbach et al. \cite{auerbach_1987} where they proved that UPOs are  experimentally accessible and capable of unfolding the structure of chaotic trajectories. In fact, by extracting the complete set of UPOs of \textcolor{black}{symbolic length up to period} $n$ and calculating their instability, they approximated the fractal dimension and topological entropy of the strange attractor of the paradigmadic H\'enon map with very high accuracy. %Cvitanovi\'c \cite{cvitanovic_1988} argued that UPOs are the optimal practical tool for measuring the invariant properties of a dynamical system and provided a solid ground for the applications of periodic orbit theory (POT). 
Artuso et al. tested this procedure through a series of applications \cite{artuso_1990, artuso_1990_2} and demonstrated that cycle expansion of the dynamical $\zeta$ function is instrumental for the analysis of deterministic chaos, even in more generic settings than the ones required by \cite{cvitanovic_1988}, i.e. when the system is not uniformly hyperbolic. % nor the results depend on the assumption of the existence of invariant measures or structural stability of the dynamics. 
Eckhardt and Ott \cite{eckhardt_1994} presented one of the first numerical applications of the periodic orbit formalism for studying the statistical and the dynamical properties of the Lorenz 1963 (L63) system \cite{lorenz_1963}. A subsequent analysis of the linear and nonlinear response of the L63 to perturbations show that specific UPOs are responsible for resonance mechanisms leading to an amplified response \cite{lucarini2009}. 

%\subsection{Application to higher dimensional systems}
%\label{higherdimsyst}
% UPOs as a proxy for turbulence
Later on, periodic orbit theory found fruitful applications also within the context of higher dimensional NESSs, and specifically in the case of (geophysical) fluid dynamics. %UPOs can be considered as a mean to simplify and interpret qualitative behaviour of a complex system \cite{lucarini_2020}, allowing to extract information and distill its dynamical structure.  This observation, together with the study of the stability and thus predictability properties of the tangent space, allows to associate relevant dynamical features of the flow to specific UPOs or classes of UPOs. UPOs, true nonlinear modes of flow, can be interpreted as a generalisation of the normal modes observed in a network of coupled linear oscillators, that allow for a study of the system in its complexity, without the necessity of considering a heavily truncated model. 
Even though a complete UPOs-based analysis of turbulent flows is still a far reaching goal, many steps have been made in this direction \cite{cvitanovic_2013}. Kawahara and Kida \cite{kawahara_2001}, who found a UPO embedded in the attractor of a numerical simulation of plane Couette flow, showed that one UPO only manages to capture in a surprisingly accurate way the turbulence statistics. \textcolor{black}{At a moderate Reynolds number}, Chandler and Kerswell \cite{chandler_2013} identified $50$ UPOs of a turbulent fluid and used them to reproduce the energy and dissipation probability density functions of the system as dynamical averages over the orbit. These encouraging results suggested that periodic orbit theory could represent a valid investigation tool also in the realm of climate systems.

\textcolor{black}{In the geophysical context, Gritsun \cite{gritsun_2008,gritsun_2013} proposed using an expansion over UPOs to reconstruct the statistics of a simple atmospheric model based on the barotropic vorticity equation of the sphere. Gritsun and Lucarini \cite{gritsun_2017}  used the UPOs for interpreting non trivial resonant responses to forcing that underlined the  violation of the standard fluctuation-dissipation relation for NESS for deterministic chaotic systems. Lucarini and Gritsun \cite{lucarini_2020} used UPOs for clarifying the nature of blocking events in a baroclinic model of the atmosphere. Specifically, they found that blocked states are associated with conditions of higher instability of the atmosphere, in agreement with a separate line of evidence \cite{Faranda2017}. Additionally, the analysis of UPOs was instrumental in proving that the atmospheric model was characterised by variability in the number of unstable dimensions, hence being not uniformly hyperbolic \cite{Lai1997}.} 

\textcolor{black}{The analysis by Lucarini and Gristun \cite{lucarini_2020} proposed the idea that the observed blocked states of the atmospheric flow should be interpreted as conditions where there is not only proximity of the \textcolor{black}{trajectory} to special classes of UPOs, but also co-evolution, at least locally in time (the so-called \textit{shadowing}). This implies that blocking can be associated with actual nonlinear modes of the atmosphere.} 

\textcolor{black}{This calls for looking at both the proximity and the co-evolution of chaotic trajectories with approximating UPOs. Recent investigations have been carried out exactly in this direction, yet in a different context. Both Yaln{\i}z and Budanur \cite{yalniz_2020} and Krygier et al. \cite{krygier_2021} investigated the process of shadowing of time-periodic solutions in three-dimensional fluids, altough using different shadowing metrics, providing a numerical evidence of the shadowing of a trajectory in terms of UPOs. In particular in \cite{yalniz_2020} the authors explored a topological approach that makes use of persistence analysis to quantify the shape similarity of chaotic trajectory segments and periodic orbits. In \cite{krygier_2021} the authors investigated whether three-dimensional turbulent flows shadow time periodic solutions. It is worth noticing that both studies investigate the properties of non-hyperbolic chaotic systems whereas the Lorenz system is almost-everywhere uniformly hyperbolic.} 

%\textcolor{black}{This calls for investigating both the proximity and the co-evolution of chaotic trajectories with approximating UPOs. Recent investigations  have been carried out exactly in this direction, yet in a different context. Yalniz and Budanur \cite{yalniz_2020} explored a topological approach that makes use of persistence analysis to quantify the shape similarity of chaotic trajectory segments and periodic orbits. Krygier et al. \cite{krygier_2021} investigated whether three dimensional turbulent flows shadow exact coherent structures.} 

\subsection{This paper}
\label{thispaper}
This paper aims at contributing to the understanding of how UPOs can be used for distilling the dynamical and statistical properties of chaotic systems. % towards the computational problems  in the application of periodic orbits theory. We believe that POT could indeed represent a very powerful tool allowing to gain insight and simplify complex systems such as the climate, but its implementation is often hindered by the numerical difficulties associated with UPOs search \cite{gritsun_2008}.
We consider the L63 model as a test case. The use of UPOs for performing accurate estimates of statistical averaging of test observables has already been extensively debated in the literature (See discussion in section \ref{model}) . 
%% parte nuova
\textcolor{black}{We will not delve into this matter, but we rather focus on shedding light on the shadowing \textcolor{black}{process}. % Specifically, we study how a long forward trajectory is approximated by a set of UPOs. We investigate both the proximity and the co-evolution of such orbit with approximating UPOs. 
Namely, at each point in time we rank in different tiers the UPOs of our database based on their distance with respect to the \textcolor{black}{trajectory} (the first tier containing the closest orbits, \textcolor{black}{the $Kth$ tier containing the $K$ closest orbits, etc.}) and we study the persistence of the ranking. Our goal is twofold. On the one hand, we aim to numerically understand how chaotic trajectories are approximated in terms of UPOs.  We anticipate that it emerges that longer period UPOs play a major role in reproducing the invariant measure of the system.  On the other hand, we study the statistics of the scattering of the orbit between the various UPOs. }

\textcolor{black}{This study of scattering uses a partition of the phase space of the L63 model that is different than the classical Ulam's partition \cite{ulam_2004}.} Each UPO (and its immediate neighbourhood) is interpreted as a  building block of the system, a spatially extended state, and the scattering can be seen  as subsequent transitions between different states; see also the recent study of a turbulent flow performed along these lines \cite{Hof2021}.

%In the second part of this paper, inspired by the work of Froyland and Padberg on quasi-invariant sets \cite{froyland_2009,froyland_2005}, we assume a global point of view on the dynamics by extracting the dynamical behaviour of the system via a Markov model \cite{froyland_2001}. 

\textcolor{black}{We will show that this viewpoint  allows for a different interpretation of quasi-invariant sets \cite{froyland_2005}. Namely, by studying the spectral properties of the discretised transfer operator, we obtain a partition of the phase space in different bundles of UPOs, each one identifying a quasi-invariant region.  We prove that UPOs represent a valid tool to investigate diffusion properties of the system, in fact, being exact solutions, they retain a memory of the geometrical structure of the attractor.}

\textcolor{black}{The structure of the rest of the paper is as follows. In section \ref{shadowing} we present the UPOs database we consider and describe our analysis of the shadowing and discuss its statistical properties.  We prove the robustness of the results independent of the shadowing criteria. In section \ref{transitionss} we construct the discretised transfer operator in terms of a finite-state stochastic matrix and use it to describe the scattering of the chaotic trajectory by the various UPOs.  We identify quasi-invariant sets through the study of the spectrum of the transition matrix and investigate the decay of correlations associated with the relaxation process of arbitrary ensemble to the invariant measure. In section \ref{conclusion} we outline our conclusions and perspectives for future works. In  Appendix \ref{newton} we provide a more extensive description of the algorithms considered for our analysis, and in Appendix \ref{quasi} we briefly recapitulate some of the main properties of quasi-invariant sets. The supplementary material provides the raw data produced in the course of this work, extra figures, videos, and further details on the methodology.\footnote{{\color{black}The supplementary material can be accessed at \texttt{https://tinyurl.com/4z6hh9a3}}.}}

\section{Shadowing of the Model Trajectory by Unstable Periodic Orbits}
\label{shadowing}
 %\subsection{Experimental Setting}
%\label{expsetting}
%In this section we numerically investigate the shadowing mechanism, with the aim of understanding whether it is possible to identify a set of key UPOs that capture the qualitative behaviour of the system. %\\
\subsection{Mathematical Framework}\label{mathframework}
We consider a continuous-time autonomous dynamical system $\dot{x}={f}({x})$ on a compact manifold $\mathcal{M} \subset \mathbb{R}^n$. We have that ${x}(t) = S^t{x_0}$, where ${x_0} = {x}(0)$ is the initial condition and $S^t$ is the evolution operator defined for $t \in  \mathbb{R}_{t>0}$. 
{\color{black}We assume that the system is dissipative (${\nabla} \cdot {f}<0$)}. We define $\Omega \subset M$ as the compact attracting invariant set of the dynamical system that supports a unique invariant physical measure $\rho$. Hence, for any sufficiently regular function (observable) $\varphi:M \rightarrow \mathbb{R}$, we have that: %  is also given, and it specifies the probability measure $\rho$, invariant and ergodic with respect  to $S^t$, in the following manner:
\begin{equation}
\langle \varphi \rangle = \int \rho(dx) \varphi (x) = \lim_{T\rightarrow \infty} \frac{1}{T} \int_0^T \varphi (S^t x_0) dt
\end{equation}
for almost all initial conditions $x_0$ belonging to the basin of attraction of $\Omega$. Another key concept we already mentioned is the one of periodic orbit. A periodic orbit of period $T$ is defined as 
\begin{equation}
\label{periodicity}
S^T(x)=x.
\end{equation}
This representation is not unique. In fact, if equation \ref{periodicity} is satisfied, $S^{nT}(x)=x$ is verified as well $\forall n \in \mathbb{N}$ . By the semigroup property of the evolution operator, we also have that $S^T(y)=y$ if $y = S^s(x)$ for any choice of $s$. From now onward we will considered a periodic orbit to be identified by its prime period $T>0$ {\color{black}(we do not consider equilibria)} and an initial condition $x_0$. 

{\color{black} We consider here \textit{chaotic} dynamical systems. By chaotic we indicate the property of sensitive dependence on initial conditions on the attractor. In particular, the first Lyapunov exponent $\Lambda_1$, which gives information on the average asymptotic rate of divergence of initially infinitesimally nearby trajectories, is positive \cite{pikovsky_2016}.

\textcolor{black}{As discussed above, the attractor of a chaotic system is densely populated by UPOs, which provide key information on the system despite being non-chaotic themselves.} Indeed, a forward trajectory on the attractor can alternatively be seen as undergoing a process of scattering between the neighbourhood of the various UPOs. For a while, the trajectory shadows - see later discussion - a nearby UPO before being repelled. The UPOs act as scattering centers exactly as a result of their instability.     
% It is known that a good understanding of the UPOs of the model plays a fundamental role in the characterisation of the system. 
Additionally, the invariant measure can be reconstructed through the use of trace formulas \cite{cvitanovic_2005} by considering the following expression for the average of any measurable observable $\varphi$: 
   
\begin{align}
\label{trace_formula}
\langle\varphi\rangle = \lim_{t \to \infty} \frac{\sum_{U^p,p\leq t} w^{U^p}\bar{\varphi}^{U^p}}{\sum_{U^p,p\leq t} w^{U^p}}
\end{align}
  
where $U^p$ is a UPO of prime period $p$, $w^{U^p}$ is its weight and $\bar{\varphi}^{U^p}$ is the average \textcolor{black}{in time} of the observable along the orbit. 
For uniformly hyperbolic dynamical systems this result is exact and the weight can be obtained, to a first approximation, by $w^{U^p}\propto \exp(-ph_{ks}^{U^p})$ \cite{grebogi_1988} , with $h_{ks}$ being the Kolmogorov-Sinai entropy of the system. This quantity provides information on the rate of creation of information due to the chaoticity of the system. From the knowledge of the spectrum Lyapunov exponents of the system $\Lambda_i$ \cite{pikovsky_2016}, %, which describe how linearised neighborhoods of the trajectory evolve in time, 
we can find an explicit expression for $h_{ks}$ via Pesin theorem \cite{ott_2002} 
\begin{equation}\\
h_{ks}\leq\sum_{\Lambda_i>0}\Lambda_i,
\end{equation}
where the left and right hand sides are equal if the invariant measure is of the Sinai-Ruelle-Bown (SRB) type \cite{eckmann_1985}.}

\subsection{The Model}\label{model}
Our analysis is performed on the L63 model, which arguably is  the most paradigmatic continuous-time chaotic systems.  The evolution equations of the L63 model are:
\begin{align*}
\dot{x}=-\sigma (x + y)\\
\dot{y} = R x - y - z x\\
\dot{z} = -\beta z + x y
\end{align*}where the three parameters $\sigma, R, \beta$ are positive numbers respectively proportional to the Prandtl number, Rayleigh number and geometry of the considered region. For specific choices of the parameters' value the attractor is a strange set and the dynamics is characterised by sensitive dependence on initial conditions  \cite{tucker_1999}. Additionally, the attractor is densely populated by an infinite number of UPOs  \cite{galias_1998}.

In this work we consider the standard parameters value $\sigma=10$, \textcolor{black}{$R=28$} and $\beta=8/3$. For such values, the dynamics of the system is characterised by a chaotic behaviour on a singularly hyperbolic attractor that supports an SRB measure \cite{tucker_2002}. %With \textit{chaotic} we indicate the property of sensitive dependence on initial conditions of the model. In particular, the first Lyapunov exponent, that gives information on the asymptotic rate of divergence of the trajectories, is positive \cite{pikovsky_2016}.\\

Many studies on UPOs of the Lorenz system have been carried out. 
\begin{comment}
 It is known that a good understanding of the UPOs of the model plays a fundamental role in the characterisation of the system. In fact, the invariant measure of the system can be reconstructed through the use of trace formulas \cite{cvitanovic_2005} by considering the following expression for the average of any measurable observable $\varphi$: 
   
\begin{align}
\label{trace_formula}
\langle\varphi\rangle = \lim_{t \to \infty} \frac{\sum_{U^p,p\leq t} w^{U^p}\bar{\varphi}^{U^p}}{\sum_{U^p,p\leq t} w^{U^p}}
\end{align}
  
where $U^p$ is a UPO of prime period $p$, $w^{U^p}$ is its weight and $\bar{\varphi}^{U^p}$ is the average \textcolor{black}{in time} of the observable along the orbit. 
For uniformly hyperbolic dynamical systems this result is exact and the weight can be obtained, to a first approximation, by $w^{U^p}\propto \exp(-ph_{ks}^{U^p})$ \cite{grebogi_1988} , with $h_{ks}$ being the Kolmogorov-Sinai entropy of the system. This quantity provides information on the rate of creation of information due to the chaoticity of the system. From the knowledge of the Lyapunov exponents of the system $\Lambda_i$ \cite{pikovsky_2016}, which describe how linearised neighborhoods of the trajectory evolve in time, we can recollect an explicit expression for $h_{ks}$ via Pesin theorem \cite{ott_2002} 
\begin{equation}\\
h_{ks}\leq\sum_{\Lambda_i>0}\Lambda_i,
\end{equation}
where the left and right hand sides are equal if the invariant measure is of the SRB type 
\end{comment}
%(as in the case of L63 for the chosen parameters' value). 
Eckhardt and Ott \cite{eckhardt_1994} presented one of the first numerical applications of the periodic orbit formalism by considering an approximate symbolic coding \cite{cvitanovic_1988} (UPOs with period up to $9$) to calculate Hausdorff dimensions and Lyapunov exponents. Franceschini, Giberti and Zheng \cite{franceschini_1993} calculated a number of UPOs of the Lorenz attractor at both standard and non standard parameter values and used them to approximate the topological entropy and Hausdorff dimension. Zoldi \cite{zoldi_1998} investigated to what extent trace formulas can can predict the structure of the histogram of chaotic time series data extracted from the run of the L63 model with different parameter values. The use of a correct weighting in the trace formula has been extensively investigated\cite{saiki_2010,saiki_2009,zaks_2010}. 

%In this paper we do not want to improve the way UPOs are used to compute sample averages, nor enlarging the already extensive database UPOs of the system, instead we wish to investigate how well a subset of UPOs can be used to reconstruct a long chaotic trajectory and interpret the transitions from one neighborhood of the orbit to another. 

\subsection{The Database}
\label{database}
\begin{figure}[h]
\includegraphics[width=7cm]{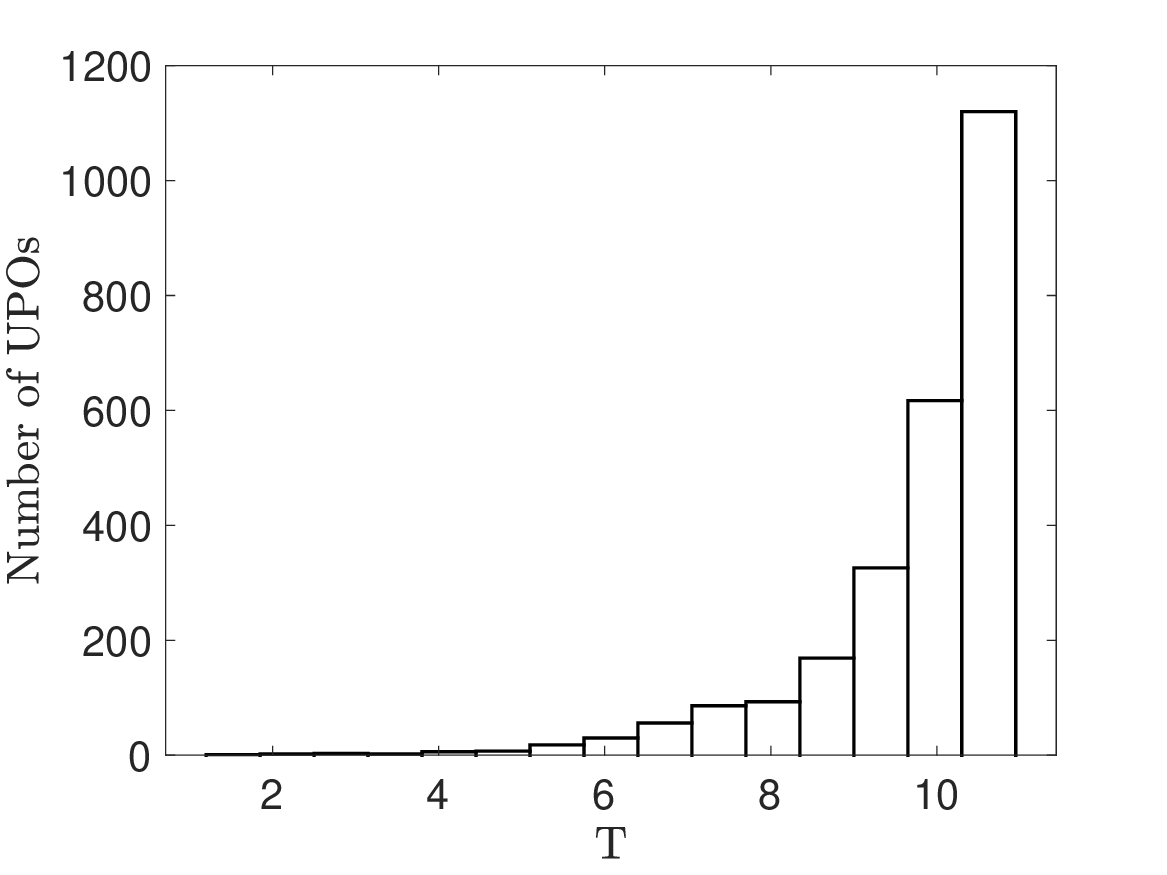}
\caption{\label{period_istogram}Number of UPOs \textcolor{black}{in our database} vs their prime period. {\color{black}We have considered symbolic sequences of period up to 14.}}
\end{figure}

%In this section we describe some general characteristics of the collected set of UPOs. 
Many numerical algorithms have been proposed so far. Saiki \cite{saiki_2007} reviewed the Newton-Raphson-Mees method, proposing a value for the damping coefficient related to the stability exponent of the orbit, while Barrio et al. \cite{barrio_2015} carried out an extensive high-precision numerical simulation in order to gather a benchmark database of UPOs for L63.
\textcolor{black}{It is possible to construct a symbolic dynamics that characterises uniquely the UPOs of the L63 model \cite{viswanath_2003}.  Motivated by the work of Galias and Tucker \cite{galias_2009}, who computed all $M=2536$ UPOs of symbolic sequence period up to 14, we use this set of UPOs for the rest of our analysis. The UPOs are computed using the Newton's method (see Appendix \ref{newton} for more details). The statistics of prime periods is shown in Fig. \ref{period_istogram}. The periods span from $T_{min}=1.5587$ to $T_{max}=10.8701$, and our sample presents the characteristic exponential growth with the period \cite{bowen_1970}. \textcolor{black}{The values of $\Lambda_1$ ranges from 0.756 to 0.994 and agree within an error of $1\%$ with the values of $\Lambda_1$ obtained in \cite{viswanath_2003}.} No UPO has a vanishing or negative value of $\Lambda_1$ (which would go against the chaotic nature of the flow).} 

%\textcolor{black}{For our analysis we decided to consider all the periodic orbits with symbolic period $T\leq 14$.Via Newton's method (see Appendix \ref{app} for more details) we obtain the exact set of $M =2536$ orbits, whose statistics of prime periods is shown in Fig. \ref{period_istogram}. The periods span from $T_{min}=1.5587$ to $T_{max}=10.8701$, and our sample presents the characteristic exponential growth with the period \cite{bowen_1970}.}
      
%\textcolor{black}{L63, being a chaotic system for the current choice of parameters' value, is characterised by the presence of a positive  Lyapunov exponent in its SRB measure, e.g. $\Lambda_1>0$ \cite{pikovsky_2016}. The distribution of values of $\Lambda_1>0$  among the UPOs included in the database ranges from 0.756 to 0.994, calculated as per \cite{benettin_1980} and agrees within an error of $1\%$ with the results obtained in \cite{viswanath_2003}.} 
Note that, as well known, the local instability of the L63 model varies wildly within its attractor, where regions with very high instability alternate with regions where one observes return-of-skill for finite-time forecast \cite{Palmer1993}. Hence in this case, as opposed to what observed in \cite{lucarini_2020}, the heterogeneity of the attractor in terms of instability cannot be explained using the properties of the individual UPOs, possibly because we are considering here a very low-dimensional flow, whereas a higher level of detail at spatial level would be needed.

%\begin{figure}[h]
%\includegraphics[width=7cm]{./lyap.eps}
%\caption{\label{lyap}Distribution of the maximal Lyapunov exponent}
%\end{figure}

\subsection{\textcolor{black}{Ranked} Shadowing of the Chaotic Trajectory}
\label{algorithm}
\textcolor{black}{We present here our results on how the UPOs rank shadow a long chaotic trajectory. The data reported below refer to a chaotic trajectory $\mathcal{X}_{chaotic}$ of duration $T_{max}=10^5$ where the output is given every $dt=0.01$. This leads to considering  the set of points $\mathcal{X}_{chaotic} = \{  x_t \}_{t=1}^{N_{max}}$  where  $N_{max}$ is $T_{max}/dt=10^7$. 
Since the system is ergodic and we consider a long trajectory compared to the timescale of the system, the statistics presented here \textcolor{black}{are} extremely insensitive to the chosen initial condition. In fact, we have repeated the same procedure for a total of \textcolor{black}{five} different chaotic trajectories of duration $T_{max}=10^5$ and all the numbers reported below oscillates of at most $1\%$, while in most cases the oscillation is only of order $0.1\%$.}  %= N_{chaotic}*dt$ , being $N_{chaotic}$ 
 % the number of points of the chaotic trajectory.

  Let us denote the set of UPOs of the database as $\mathcal{U}=\{U_k\}_{k=1}^{M}$ where the UPO $U_k$ is intended as a set of points in the system phase space $U_k=\{ u_k(s)\} _{s=1}^{dt*T_k}$, with $T_k$ being its period and $dt$ the time step. %We consider a numerical chaotic trajectory $\mathcal{X}_{chaotic}$ consisting of the set of points $\mathcal{X}_{chaotic} = \{  x_t \}_{t=1}^{N_{max}}$  where  $N_{max}$ is %= N_{chaotic}*dt$ , being $N_{chaotic}$ 
  the number of points of the chaotic trajectory. 
\textcolor{black}{  
We define a metric of proximity that allows us to select and rank the closest UPOs to the trajectory at each point in time. More precisely, we say that the UPO $U_{\bar{k}}$ has the closest pass to the chaotic trajectory $\mathcal{X}_{chaotic}$ at time t if}
    
\begin{align}
\label{shadowing_definition}
\min_s| u_{\bar{k}}(s)-x(t)| = \min_k(\min_s|u_k(s)-x(t)|)
\end{align}
  
\textcolor{black}{It is important to notice that closeness and shadowing become equivalent when the distance given in Eq. \ref{shadowing_definition} becomes infinitesimal. The minimal distance between a UPO and the chaotic trajectory decreases  as we consider complete sets of UPOs with larger and larger maximum symbolic length. The statistics of such distance for the case studied here is shown in Fig. \ref{distance} and discussed below. We can then define the \textit{ranked shadowing}, where for each point $x_t$ along the chaotic trajectory $\mathcal{X}_t$ we rank the UPOs according to their distance from $x_t$. Note that after a time step the distance between a given UPO and the chaotic trajectory will change, while its rank UPO might stay the same or also change. The supplementary material includes a hopefully informative video that illustrates how UPOs shadow the chaotic trajectory.}

%select a set of $N$ ranked UPOs consisting of the first $N$ UPOs that minimise the distance with $x_t$. In this fashion, at each time step $t$ the UPOs are ranked in $N$ tiers, accordingly to their distance with the chaotic trajectory.}

This calculation was carried out using all available periodic orbits, using an output time-step $dt=0.01$ (See Appendix \ref{newton} for more details on the algorithm specification). Clearly, it is important to test whether all the UPOs of our database rank shadow at least once the chaotic trajectory.

\begin{figure}[h]
\includegraphics[width=7cm]{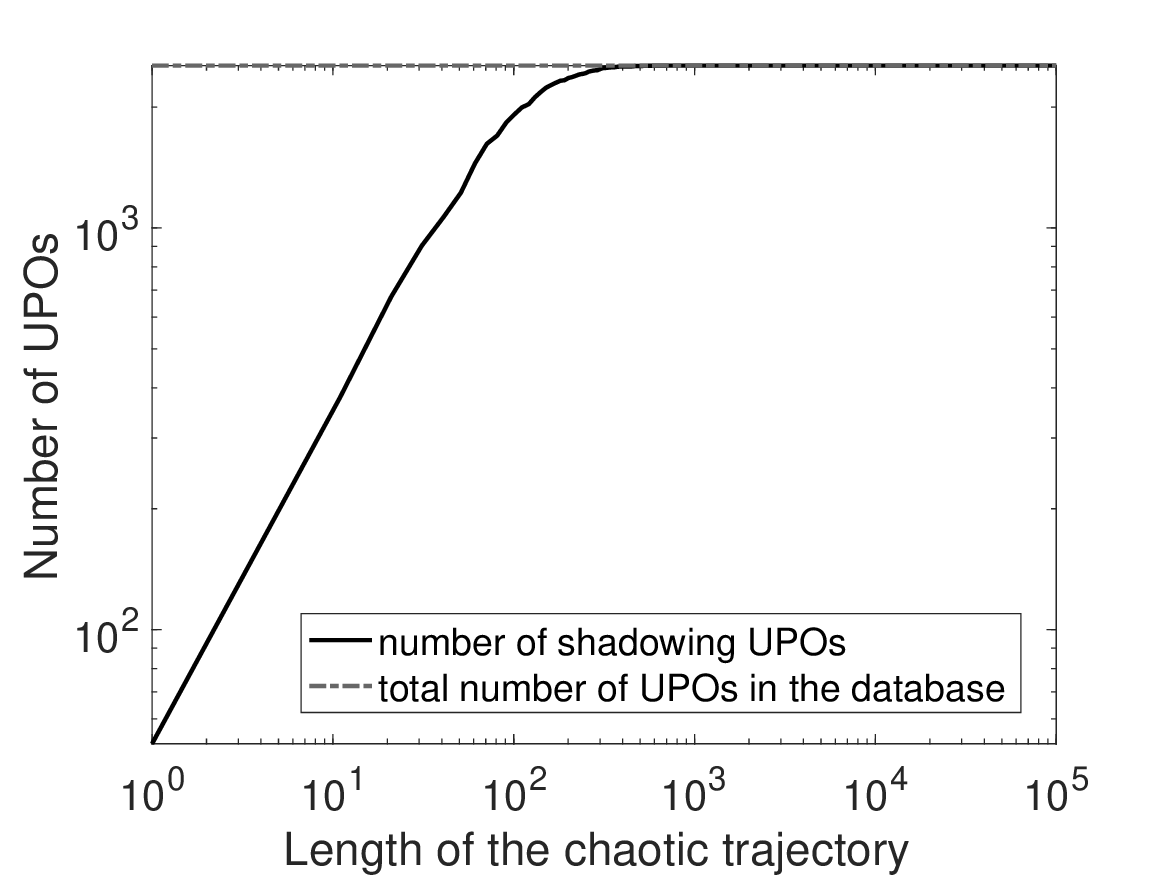}
\caption{\label{num_UPO}Number of shadowing UPOs as a function of the length of the shadowed chaotic trajectory}
\end{figure}

\textcolor{black}{We can see in fact from Fig. \ref{num_UPO} that the number of UPOs $N_U(t)$ that perform rank shadowing at least once grows very rapidly with the length of the trajectory $t$. We find an approximate power law $N_U(t)\propto t^{\alpha}$ with $\alpha \approx 0.78$ for moderate values of $t$ up to $\approx 100$. %the relation between the length of the chaotic trajectory and the number of UPOs that perform shadowing at least once can be approximated with a power law $\propto x^{\alpha}$ with $\alpha \approx 1.28$. 
A chaotic trajectory having a duration of $10^3$ time units already saturates the database, so that when considering a trajectory of duration $T_{max}=10^5$ all UPOs in the dataset shadow the trajectory multiple times.} %We believe that this is the ideal length to produce reliable statistics and from now onward all the statistics produced are derived from data obtained by shadowing a chaotic trajectory of length $T_{max}=10^5$ with the full database.

\begin{figure}
    \subfloat[]{%
      \includegraphics[width=7cm]{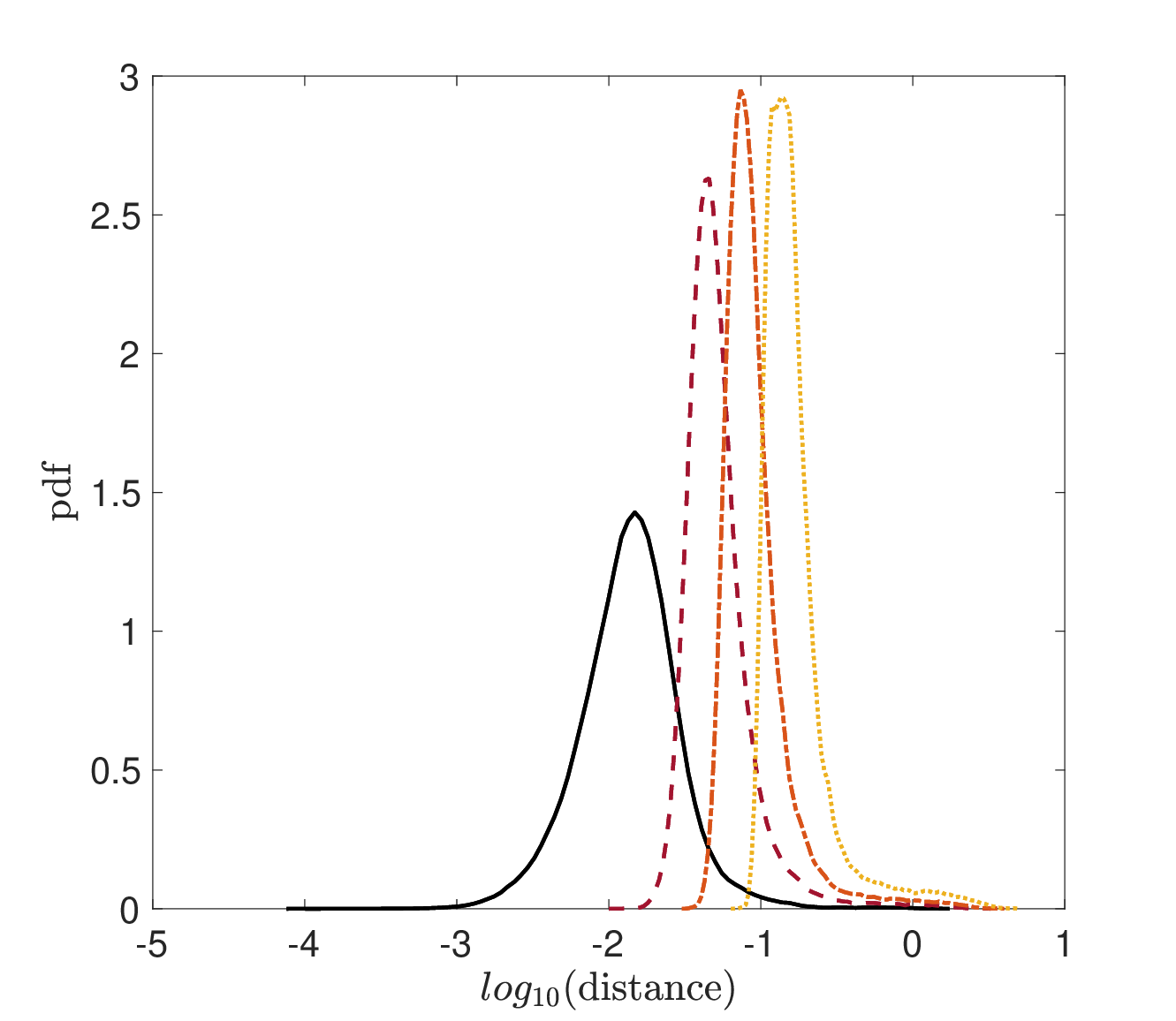}
           \label{distance}
    }
    \qquad \qquad
     \subfloat[]{%
   \includegraphics[width=7cm]{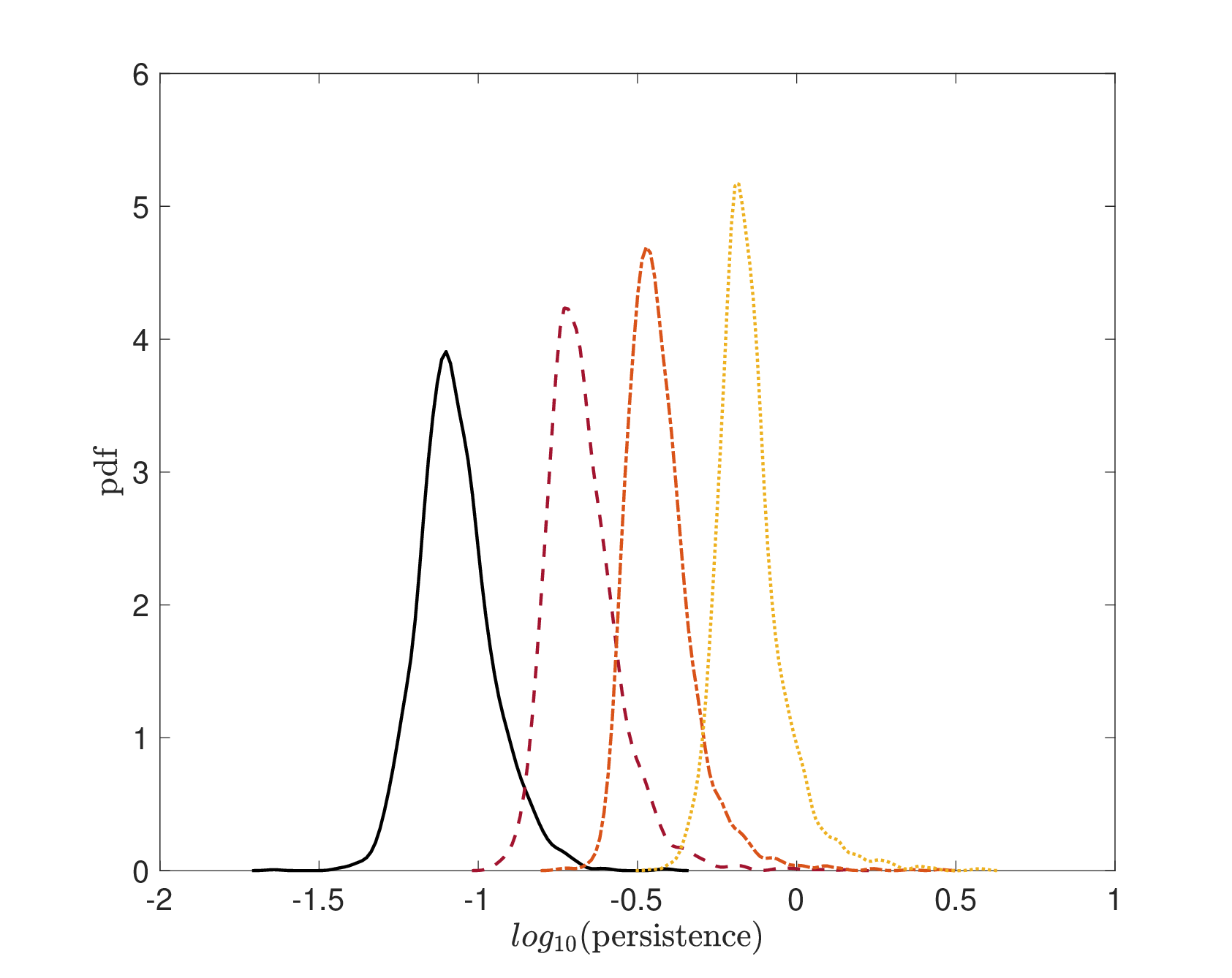}
              \label{persistence}
    }
       \caption{\color{black}{Panel \protect\subref{error_bar_shadowing}: Probability distribution function for the $log_{10}$-distance distribution of the first tier orbits (solid black  line; mean distance  0.0189), tier $K=10$ orbits (dashed red line; mean distance 0.0649 dashed red), tier $K=30$ orbits (dashed and dotted orange line; mean distance 0.1130), and tier $K=100$ orbits (dotted yellow line; mean distance 0.2106). Panel \protect\subref{error_bar_occupancy}: Probability distribution function of $log_{10}$-persistence of the tier 1 orbits (solid black line; mean persistence 0.0880), and of the shadowing orbits with modified definition allowing for fluctuations withing the first $K=10$ tiers (dashed red line; mean persistence 0.2218), $K=30$ tiers (dashed and dotted orange line; mean persistence 0.3846), and $K=100$ tiers (dotted yellow line; mean persistence  0.7371). See the main text for further details.} }
  \label{shad} 
\end{figure}

\textcolor{black}{The reader might think that the definition of shadowing proposed in Equation \ref{shadowing_definition} could be unreasonably strict. In fact, at each time step we are only  selecting the nearest UPO, thus possibly discarding many other UPOs that are also extremely close to the trajectory. % that minimises the distance with the chaotic trajectory, namely the shadowing orbits in tier 1. 
Hence, we also propose a looser definition of shadowing that allows to take into account the fact that a UPO might still be nearby the trajectory even if it is not anymore the nearest one. In particular, if $U_t$ is the closest UPOs to the trajectory at time $t$, we say that $U_t$ persists in shadowing  at time $t+1$ if by then $U_t$ is one of the $K$ closest UPOs, or, in other terms, it belongs to one of first $K$ tiers. In this fashion we are rewarding the quality of the shadowing of the UPOs within the first $K$ tiers. When the UPOs exits the first $K$ tiers of shadowing, the closest UPO to the trajectory is selected as shadowing UPO. In this manuscript we will consider various values of $K$ ($K=1$ corresponding to the original, strictest definition of shadowing) in order to assess the robustness of our results.}

\textcolor{black}{In general, the shadowing UPOs are characterised by two properties. First, by definition, they have a close proximity  with the chaotic trajectory. Additionally, since the flow is smooth, we expect a certain degree of persistence in the shadowing: if a UPO is near the chaotic trajectory, the velocity fields will also be similar, and one expects that the UPO will persist its shadowing property for a certain time. The persistence, namely the mean time duration of the shadowing process, quantifies the temporal co-evolution of the chaotic trajectory with the approximating UPOs. In the present discrete numerical implementation of the ranked shadowing process it is possible that the closest UPO might not be the orbits that has the higher persistence. However, even in the case of existence of another orbit with higher persistence, the bounds on the velocity field and, more importantly, on the norm of the Jacobian of such field, guarantee that the selected orbit, chosen solely based on the proximity criteria, would stays close to the trajectory for a certain period of time. We could quantify this information by noticing that the mean speed over the attractor is about $26$ with stdev $9$. This results on an average displacement of about $0.26$ for the considered numerical discretisation $dt=0.01.$ }

\begin{table}[]
\label{probability}
\begin{tabular}{@{}llll@{}}
\toprule
                                & $P(d>1)$ & $P(d>10^{-1})$ & $P(d>10^{-2})$ \\ \midrule
\multicolumn{1}{l|}{$tier$ 1}   & 0.0001   & 0.0096         & 0.6891         \\
\multicolumn{1}{l|}{$tier$ 10}  & 0.0026   & 0.0816         & 1              \\
\multicolumn{1}{l|}{$tier$ 30}  & 0.0076   & 0.2997         & 1              \\
\multicolumn{1}{l|}{$tier$ 100} & 0.0230   & 0.9551         & 1              \\ \bottomrule
\end{tabular}
\caption{\label{probability}Probability that the distance between the chaotic trajectory and the shadowing UPO exceeds the indicated thresholds.}
\end{table}

{\color{black}\textcolor{black}{Fig.} \ref{distance} presents the probability distribution functions (pdfs)  of the distance of the shadowing UPOs for tiers $K\in\{1,10,30,100\}$.By definition, as we look at successive tiers, the average distance of the shadowing UPOs with the chaotic trajectory increases, going from $\mathcal{O}(10^{-2})$ for $K=1$ up to $\mathcal{O}(10^{-1})$ for $K=100$. More precisely, the mean distance is respectively 0.0189,  0.0649, 0.1130 and 0.2106 for the orbits in tier 1, 10, 30 an 100. One should keep in mind that the tier $K=100$ includes the top $4\%$ of the UPOs. Note that substantial overlaps exist between the various pdfs, thus indicating that, in absolute terms, the quality of the shadowing varies throughout the attractor. As we could further quantify in Table \ref{probability}, the quality of the shadowing is in general very high: even considering the weakest definition of shadowing, only about $2\%$ of the recorded distances are above 1. Choosing the strictest definition of shadowing, only $0.1\%$ of the recorded distances are above 0.1. This can be better appreciated also by considering that the attractor of the L63 model is contained in the Cartesian product $\mathcal{P}=[-20,20]\times [-27.5,27,5]\times [1,48]$ \cite{sparrow_1982}. One can cover this region with $103400\times10^{3l}$ cubes of equal size $10^{-l}$. We will use such a partition (for $l=0$) later in the paper.

\textcolor{black}{Figure} \ref{persistence} shows the distribution of the mean persistence of the shadowing UPOs when we consider the strict as well as looser definitions of shadowing, with $K\in\{1,10,30,100\}$. By construction, the mean persistence increases with $K$ as we are using looser and looser criteria for defining it. Note that in all cases the time persistence is strictly larger than \textcolor{black}{four} time steps, meaning that our procedure captures in all cases at least some co-evolution of the chaotic trajectory and of the approximating UPOs. This also suggests that the adopted temporal resolution for our chaotic trajectory and UPOs is sufficient: had we chosen a longer time step, we would have lost the property of co-evolution. Specifically, the mean persistence is 0.0880, 0.2218, 0.3846, 0.7371 (corresponding to approximately 9, 22, 38, and 74 time steps) when allowing for fluctuations respectively in the first and first 10, 30, and 100 tiers. In the latter, case, persistence is of the same order as the Lyapunov time ($\Lambda_1^{-1}$). These average temporal durations translate into average rectified distances of co-evolution of about 2, 5, 10 and 19. These figures are larger by a factor $\mathcal{O}(10^2)$ than the corresponding average distances between the chaotic trajectory and the shadowing UPOs, thus reinforcing our claim that the shadowing is accurate and persistent.

%If we compare the rectified distances with the numerical discretisation of the phase space we note that the co-evolution of the UPO with the trajectory crosses some, up to many boxes of the partition. This also suggests that the adopted temporal resolution for our chaotic trajectory and UPOs is sufficient: had we chosen a longer time step, we would have lost the property of co-evolution. %Interestingly, the mean persistence ranges from $\mathcal{O}(10^{-1})$ for $K=1$ up to $\mathcal{O}(1)$ in units of Lyapunov time ($\Lambda_1^{-1}$) for $K=100$. 

%Lastly, it is interesting to notice that comparing Table \ref{probability} with the distance presented in Fig. \ref{shad} we understand that the typical distance between the chaotic trajectory and the shadowing UPO is orders of magnitude smaller than the co-evolving trajectory mean distance regardless of our precision, meaning that we are indeed capturing instances of co-evolution of the trajectory with the UPO.
}

\subsection{Longer Period UPOs Shadow the Trajectory for a Longer Time}
\label{occupancy}
%Our main goal is to shed light on the shadowing mechanism.
%It is crucial to develop some tools that are capable of measuring to what extent each UPO contributes in the approximation of the chaotic trajectory.
%With this purpose in mind, we define the \textit{shadowing time} and \textit{occupancy ratio} of a UPO.
We define the \textit{shadowing time} of a UPO  as the total amount of time that the UPO spends shadowing the chaotic trajectory. More precisely, if the UPO ${U}_k$ is selected as shadowing orbit $t_k$ times, its shadowing time will be $r_k=t_k*dt$ . This quantity is a good indicator for the absolute shadowing time, but it does not take into account the length of the UPO. Longer period UPOs correspond to a longer trajectory in phase space. We then introduce the \textit{occupancy ratio} for the UPO $U_k$, defined as $o_k= \frac{t_k}{T_k/dt} $ with $T_k$ being the period of the UPO. In this way we are able to measure the shadowing time normalised over the period of the UPO. An occupancy ratio much larger than one indicates that it is likely that a large portion of the UPO has shadowed the trajectory at least once. 

\begin{figure}
    \subfloat[]{%
      \includegraphics[width=7cm]{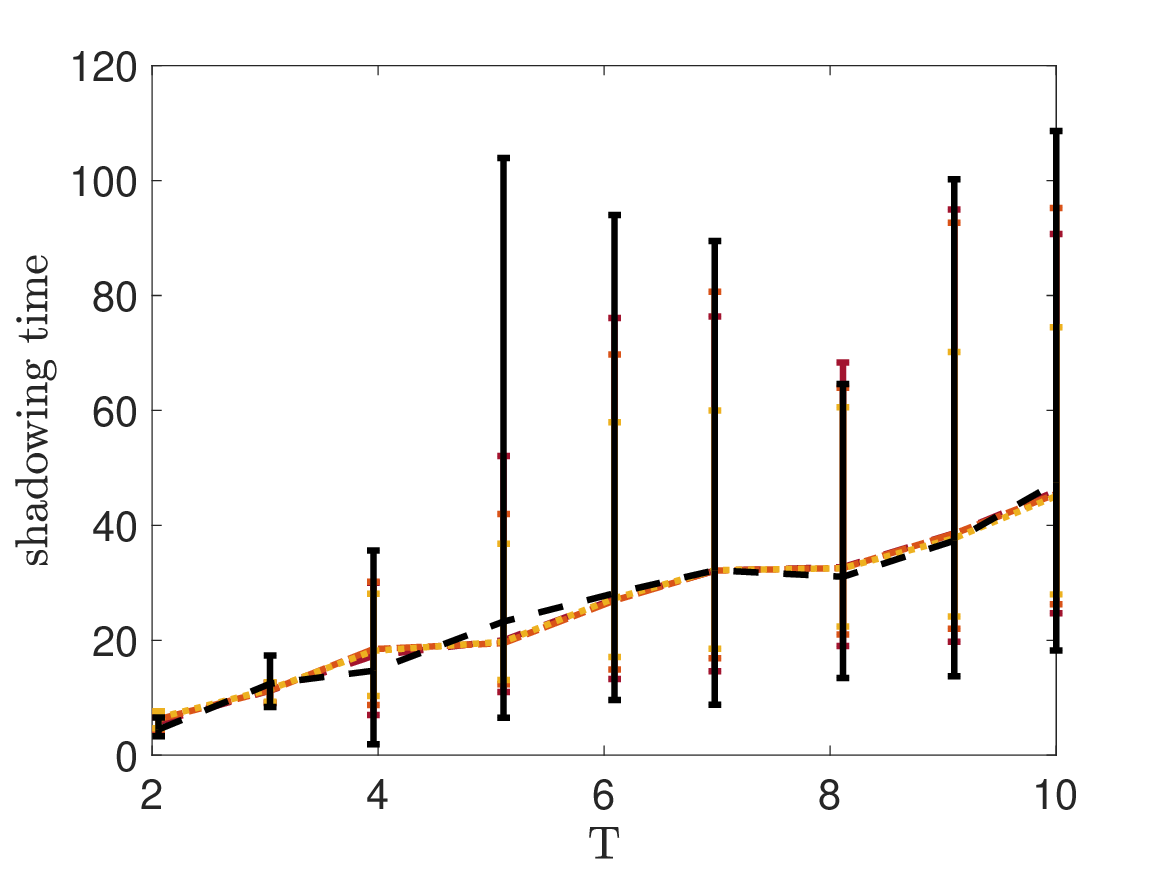}
           \label{error_bar_shadowing}
    }
    \qquad \qquad
     \subfloat[]{%
   \includegraphics[width=7cm]{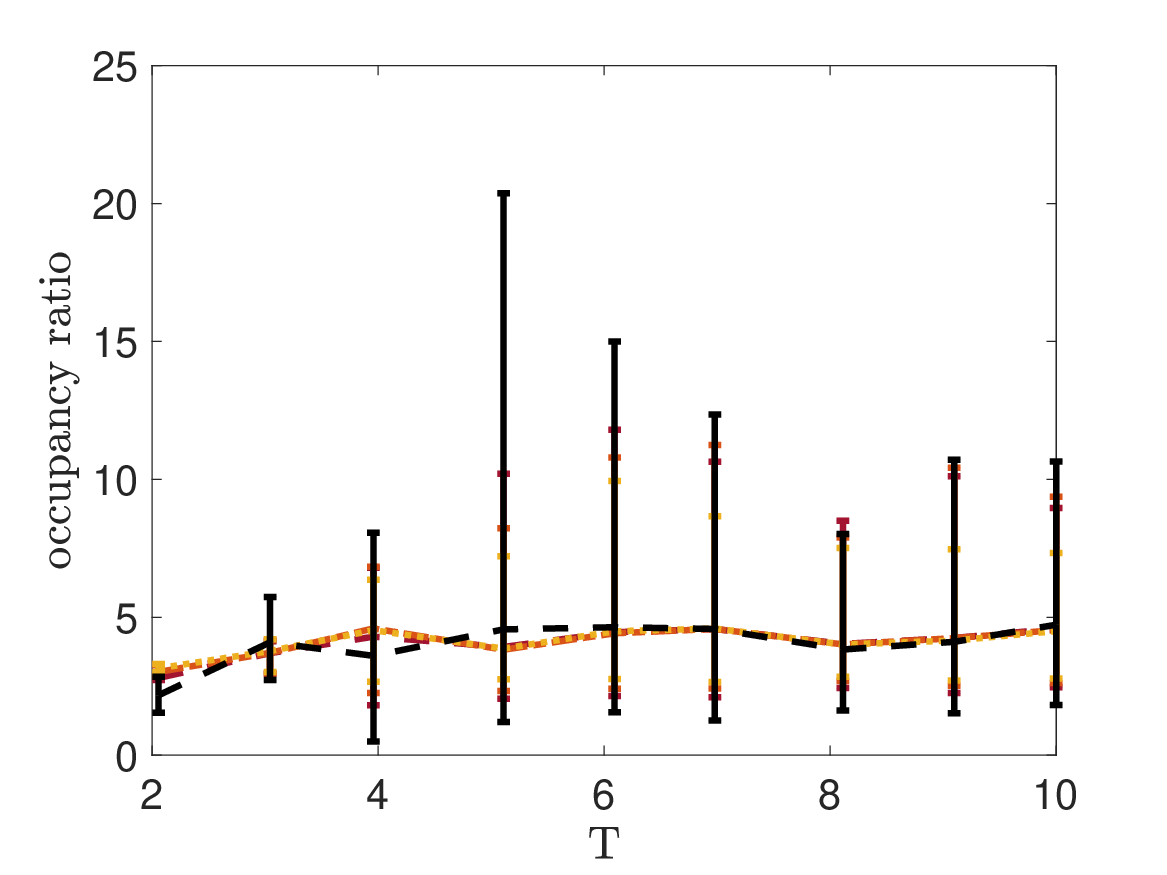}
              \label{error_bar_occupancy}
    }
       \caption{Average shadowing time (panel \protect\subref{error_bar_shadowing}) and occupancy ratio (panel \protect\subref{error_bar_occupancy}) of the first tier  (dashed black line) and averaged over first 10 (dashed red), 30  (dashed and dotted orange line) and 100 (dotted yellow line) tiers for UPOs of period T. The bars indicate the range between the percentiles $2.5$ and $97.5$ for each value of T.}
  \label{error_bar} 
\end{figure}

\begin{figure}[h]
\includegraphics[width=7cm]{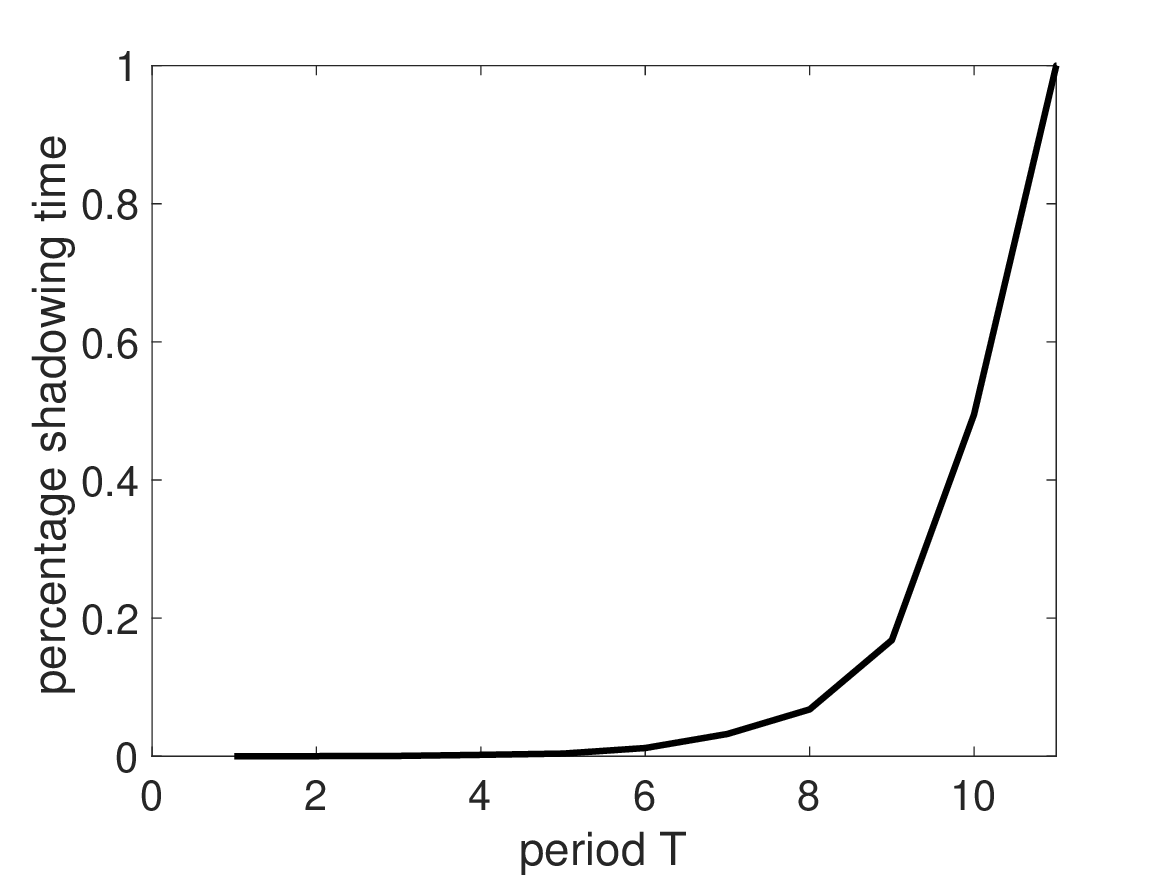}
\caption{\label{percentagetime_period.jpg}{\color{black}Cumulative fraction of the shadowing time performed by UPOs having larger and larger period.}}
  \label{cumulative} 
\end{figure}

\textcolor{black}{One could interpret the trace formula given in Eq. \ref{trace_formula} as suggesting that on the average low period orbits should dominate in terms of shadowing a chaotic trajectory, because the statistical weight of long period orbits is exponentially suppressed. Instead, %we find that higher period orbits have a major role in reproducing the invariant measure of the system. In fact, we would expect longer orbits to have a shorter shadowing time, according to the weighting given in equation \ref{trace_formula}. Instead, 
as shown in Fig. \ref{error_bar}, the shadowing time increases with the period of the UPOs, while the occupancy ratio remains the same. This means that, by and large, all the UPOs are selected to shadow the chaotic trajectory with the same weighting, independently of their period. However, since the number of periodic orbits grows exponentially with the period (see Fig. \ref{period_istogram}) longer orbits overall dominate, as shown in Fig. \ref{cumulative}.}

In order to assess the robustness of our results, we have studied the shadowing orbits in the first $K$ tiers, with the goal of testing whether even allowing for a looser definition of shadowing UPOs, the role of longer orbits remains consistently dominant. In this context, we are interested in average quantities over all tiers. Namely, we define the \textit{average occupancy ratio} at time $t$ as

\begin{align}
\bar{o}_t = \frac{1}{K}{\sum_{k=1}^K o_k}
\end{align}
where $o_k$ is the occupancy ratio of the UPO that shadows the trajectory at time t in tier $k$. \textcolor{black}{Similarly we define the\textit{ average period} and \textit{average shadowing time} at time $t$. As mentioned above, a given UPO might appear in different tiers at different times.}

%Note that in each tier the shadowing UPO is chosen among the full set of orbits, meaning that a cycle could shadow the trajectory at \textcolor{black}{different timesteps} $t$ in more than one tier.}

\label{longer}

The robustness of the analysis is confirmed when reproducing the statistics presented in Fig. \ref{error_bar} with $K$ shadowing UPOs. Allowing for more shadowing UPOs does not affect the correlation found in the previous section when considering average quantities. Note that the numbers reported in Figs. \ref{error_bar}a,b scale proportionally to $T_{max}$. %We have clear evidence that results are robust when undersampling (we are trying to check what happens if we are unable to find all the $M=2536$ UPOs) the set of UPOs by a factor up to 100. 

%They provide an insight into the shadowing dynamics and are particularly useful in the applications of periodic orbits theory. 
These findings, which seem at odds with what the trace formula seems to indicate, support the idea that long period orbits play an important role for computing ensemble averages \cite{Zoldi1998,lasagna_2018,lasagna_2020}. % and provide support to the approach proposed by Lasagna in \cite{lasagna_2018,lasagna_2020}, where it is suggested that one should focus the computational resources on finding long period UPOs. 

%In fact, knowing a priori that longer UPOs play a major role in reproducing the invariant measure of the system could give a contribution to the optimisation of periodic search algorithms by addressing computational resources to a limited subset of UPOs.
%Hence our findings seem to provide support to the approach proposed by Lasagna in \cite{lasagna_2018,lasagna_2020}, nonetheless it is not clear to us why this might hold true.
\section{TRANSITIONS}
\label{transitionss}
In this section we use UPOs as a tool to investigate the mixing properties of the system. The \textit{ranked shadowing} will be used to define the Markov process that describes the sequence of transitions between neighbourhoods of UPOs that define the time evolution of the chaotic trajectory.

  \subsection{Extracting a Markov Chain from the Dynamics}
  \label{markovchain}
 
A very valuable tool to study transitions and evolution of measures in dynamical systems is offered by the transfer operator. \textcolor{black}{On the attractor $\Omega$ the \textit{Perron-Frobenius operator} or \textit{transfer operator} $\mathcal{P}_t: L^1(\Omega)\to L^1(\Omega) $, is defined as   
\begin{align}
\mathcal{P}_t \rho(x)=\int_{\Omega} \rho(y)\delta(x-S^t(y))dy= \rho(S^{-t}(x))|det(DS^{-t}(x))|,
\end{align}
which evolves probabilities densities $\rho$ under the dynamics of the system; note that $D$ indicates the Jacobian.} 
%The transfer operator is a statistical tool that describes how the densities globally evolve with time. 
From the study of its spectral properties we can deduce significant statistical information about the system, such as mixing properties, invariant densities and decay of correlations \cite{baladi_2000,froyland_2001}. For instance, fixed points of $\mathcal{P}_t$ represent invariant densities for the dynamics, that remain unaltered by the flow. \\
% Discretised transfer operator
We need to define an appropriate numerical estimate of the transfer operator $\mathcal{P}_t$.                                                   
In fact, in order to tackle the problem from a numerical standpoint, we have to consider the transfer operator within a finite dimensional setting, where the phase space is not interpreted as a continuum, but it is appropriately discretised into a finite collection of regions, with mass moving from one region to the other at each iteration of the transfer operator. It is important to notice that at this stage the dynamics occurring within each set of the partition is ignored, and we are just interested in the macroscopic movement of mass. 
Different methods for defining this approximation have been developed. For the well-known Ulam's method \cite{ulam_2004} the approximation takes the form of a regular lattice covering the phase space. {\color{black}See \cite{Froyland1998,froyland_2001} for classical results on the use of the Ulam's method for approximating the properties of chaotic dynamical systems and \cite{lucarini2016,gutierrez_2020} for recent applications on the L63 model.}
\begin{comment}
More precisely, let $\{B_i\}_{i=1}^M$ be a partition of the phase space in $M$ boxes of identical size, so that $\Omega = \bigcup_{i=1}^M B_i$. Let $\chi_{B_i}$ be the characteristic function of $B_i \subset \Omega$. We consider the projection $\pi_M: L^1(\Omega) \to sp\{\chi_{B_1},...,\chi_{B_M}\}$ given by 
    
\begin{align}
\pi_M f = \sum_{i=1}^M \left (\frac{1}{m(B_i)}\int_{B_i} f \quad dm \right) \chi _{B_i},
\end{align}
  
with $m$ being the normalised Lebesgue measure on $\Omega$.
The action of $\mathcal{P}_t$ on $sp\{\chi_{B_1},...,\chi_{B_M}\}$ admits the following matrix representation
    
\begin{align}
\label{transitions}
P_{i,j}^{\tau}=\frac{m(B_i\cap \Phi ^{-\tau}B_j)}{m(B_i)}.
\end{align}
  
By construction we have that $P^{\tau}$ is a stochastic matrix. It follows that $P_{i,j}^{\tau}\geq 0$ and $\sum_{i=1}^M P_{i,j}^{\tau}=1$ $\forall j=1,...,M$, since $P_{i,j}^{\tau}$ gives the probability of being in state $i$ at time $t+\tau$ being the system in state $j$ at time $\tau$. 
\end{comment}

% our innovation: partition in terms of UPOs
We propose here a different way to discretise the dynamics of the system. {\color{black}Similarly to what done in \cite{Hof2021},} we select $M$ numerical UPOs $U_1,..., U_M$ and we associate the states $A_1,..., A_M$ obtained by considering the UPOs together with their neighbourhoods. Each $A_i$ represents one of the possible discrete states of the system. We implement the shadowing algorithm: at each time step $t$ the UPO $U_k$  that minimises the distance with the chaotic trajectory is selected (See section \ref{algorithm} for more details on the algorithm). Hence we say that the system is in the state $A_k$ at time $t$. The stochastic variable $s: \{1,...,N_{max}\}\subset\mathbb{N}\to \mathcal{A}$ describes the shadowing process just outlined as follows:
    
\begin{align}
s(t) = A_k
\end{align}
  
with $A_k$ being the shadowing UPO at time t and corresponding neighbourhood. 
We then construct the stochastic matrix as %along the lines of equation \ref{transitions} as 
    
\begin{align}
P_{i,j}^{dt}\approx \frac{\# \{ k: (s(k) = A_j) \wedge  (s(k+dt)=A_i)\} }  {N_{chaotic}}
\end{align}    where $\#$ defines the cardinality of the set.
\subsection{Spectral Properties of the Transfer Operator}
\label{transferoperator}
In this section we use the spectrum of the stochastic matrix $P^{dt}$ to study the mixing properties of the system. We focus on the process of scattering that the forward trajectory undergoes by being repelled continuously between the neighbourhood of the various UPOs. 

Let us recall a few basic properties of the spectrum of a general stochastic matrix. Its leading eigenvalue is $\lambda=1$, and its corresponding eigenvector $\textbf{w}^{(1)}$ , in the case of an ergodic Markov chain, determines the unique invariant measure. The other eigenvalues, which can be proven to be inside the unit circle, fulfill the condition $\sum _jw_j^{(\lambda)}=0$, where $w^{(\lambda)}_j$ indicates the $j^{th}$ component of the eigenvector $w^{(\lambda)}$. The subdominant eigenvalues , ordered accordingly to $1>\Re(\lambda_2)\geq \Re(\lambda_3)\geq ... \geq \Re(\lambda_M)$ (where $\Re$ indicates the real part) can be thought of as modes of decay, as they determine the time scale of convergence to the stationary probability measure. We can quantify these time scales by defining the corresponding decay rate as $\tau_k = -\frac{dt}{log\left(\Re\left(\lambda_k\right)\right)}$, where $dt$ takes into account how we have discretised  the dynamics in the time domain. In particular, $\tau_2$ identifies the mixing time scale  \cite{pikovsky_2016}.
%In particular, it is possible to quantify the mixing time scale as $\tau_2=-\frac{dt}{\log(\Re(\lambda^{(2)}))}$ in terms of the first subdominant eigenvalue $\lambda^{(2)}$ \cite{pikovsky_2016}.

\textcolor{black}{We derive the matrix $P^{dt}$ following the procedure outlined in Section \ref{markovchain}, by considering the shadowing of a chaotic trajectory with length $T_{max}=10^5$ with the full set of $M=2536$ UPOs. $P^{dt}$ is a stochastic matrix by construction, its first eigenvalues are $\lambda_1=1, \lambda_2 = 0.9841, \lambda_3=0.9806, \lambda_4 = 0.9706$ and the corresponding decay rates are $\tau_2= 0.6239 , \tau_3 = 0.5104, \tau_4 = 0.3351$. We also verified  that \textcolor{black}{there exists} a value $\hat{N}$ so that $P_{i,j}^{\hat{N}}\neq 0$ $\forall$ $i,j$, implying that the process is ergodic. Additionally, we tested the markovianity of the process by verifying that the stocastic matrix $P^{ndt}$ defining the scattering sampled every $n>1$ time steps of the chaotic trajectory between the neighbourhoods of the various UPOs has very similar dominant eigenvectors as those of $P^{dt}$, while the corresponding eigenvalues scale, with a good approximation, with the $n^{th}$ power, as expected.} \\

\subsection{Quasi-Invariant Sets}
\label{quasiinvariant}

\begin{figure}[h!]

        \subfloat[]{%
    
      \includegraphics[width=7cm]{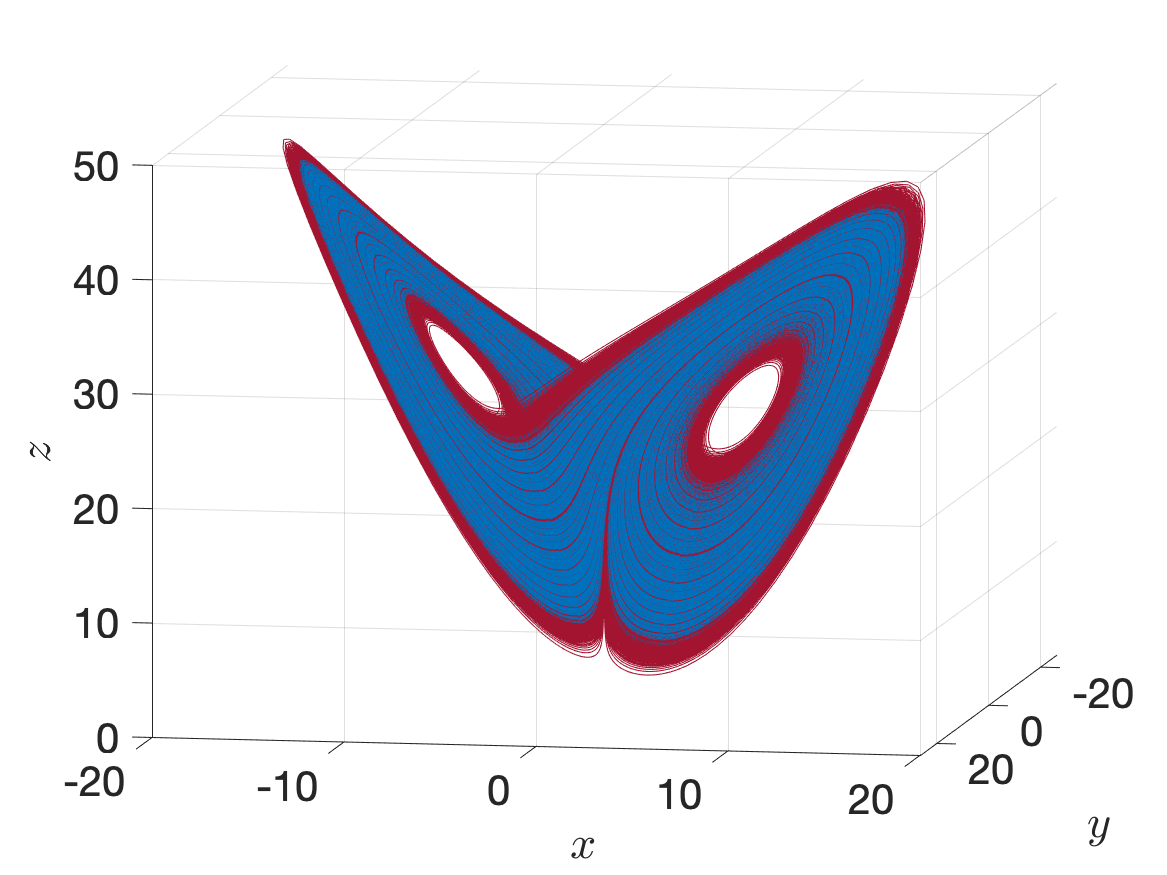}
           \label{fig: subfig 9a}
    }
    \qquad \qquad
          \subfloat[]{%
      \includegraphics[width=7cm]{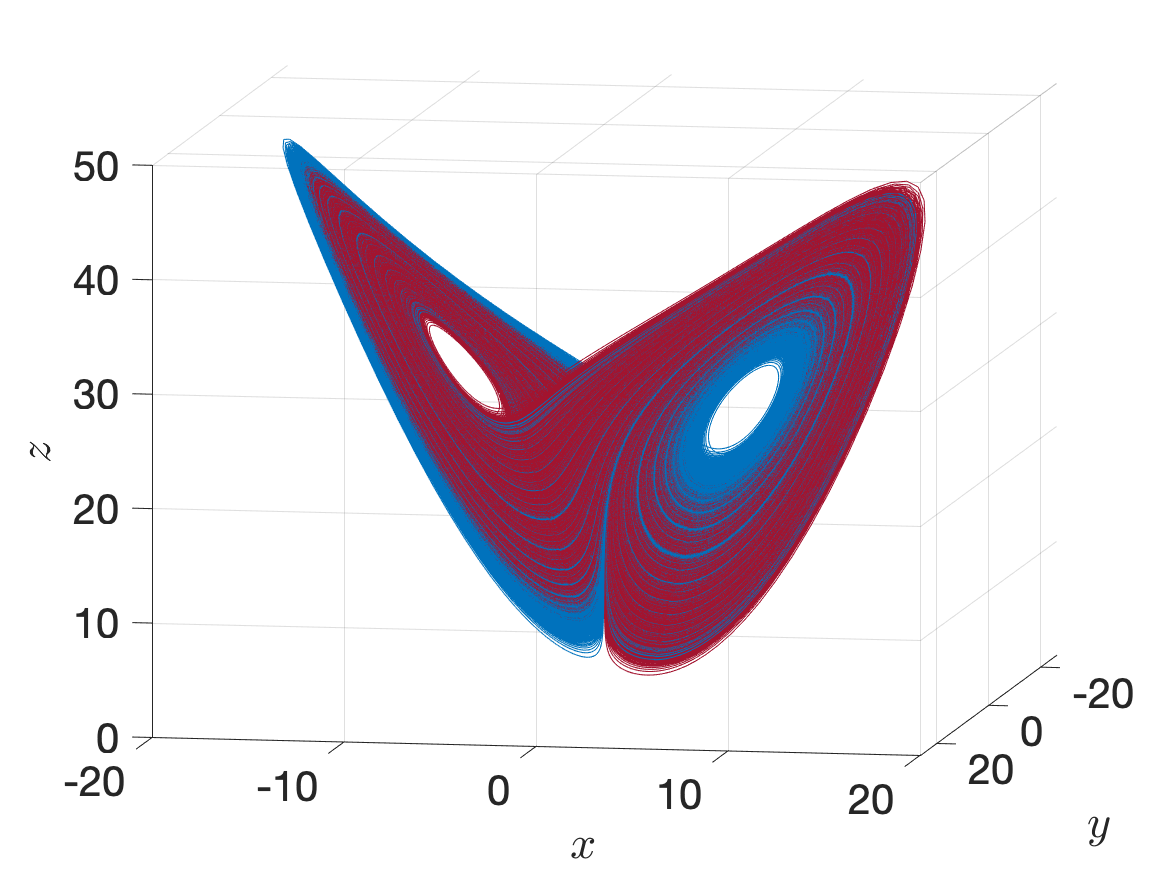}
           \label{fig: subfig 9b}
    }
    \qquad \qquad
          \subfloat[]{%
      \includegraphics[width=7cm]{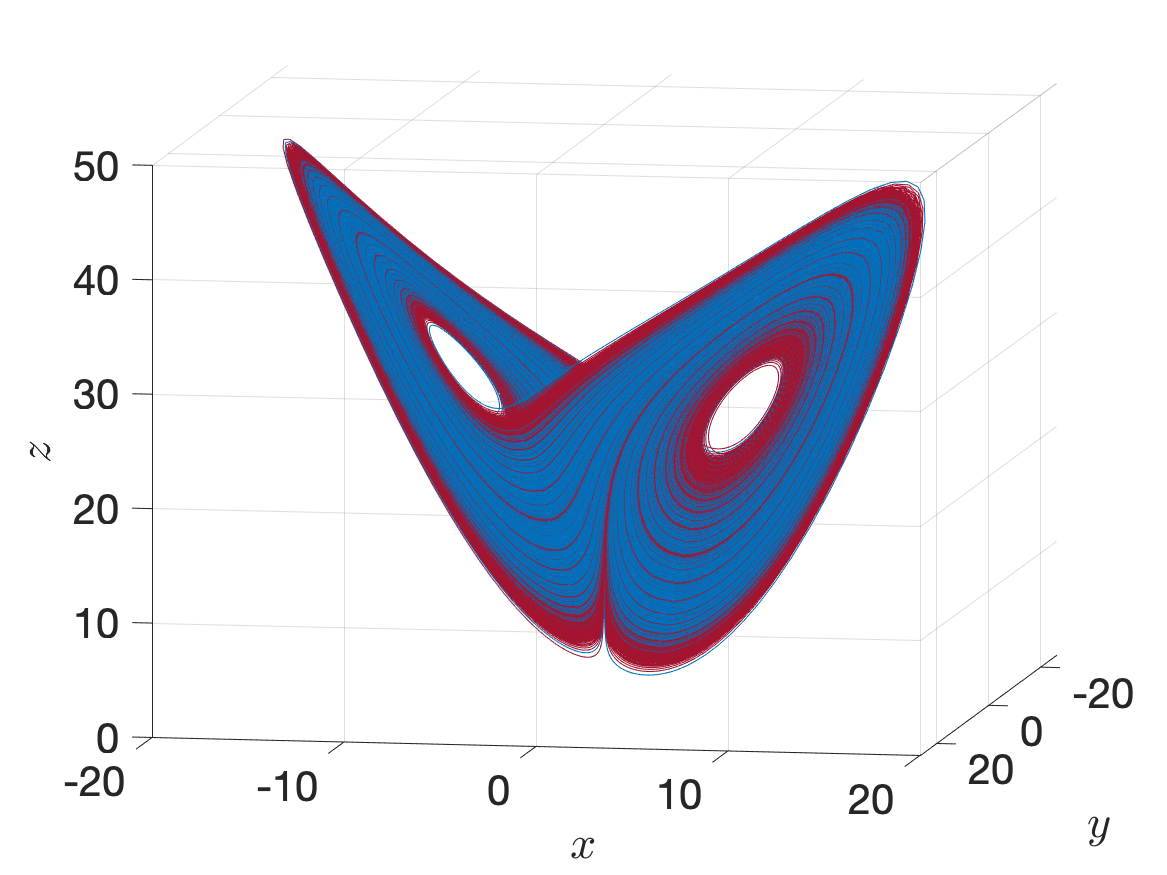}
           \label{fig: subfig 9c}
    }
    \caption{ \label{bundle} Quasi-invariant bundles of UPOs obtained with the method outlined in Section \ref{quasiinvariant}. \protect\subref{fig: subfig 9a}: $\lambda_2=0.9841$, $\tau_2 = 0.6239$;  \protect\subref{fig: subfig 9b}: $\lambda_3=0.9806$, $\tau_3 = 0.5104$; \protect\subref{fig: subfig 9c}: $\lambda_4=0.9706$, $\tau_4 = 0.3351$.}
\end{figure}

We wish to attempt an interpretation of the eigenvectors of $P^{dt}$ corresponding to the subdominant eigenvalues. Let $\textbf{w}^{(k)}$ be the eigenvector associated with $\lambda_k$, $k\geq2$. This allows us to define two sets $B_1$ and $B_2$:

\begin{align}
B_1 = \bigcup_{i \in \mathcal{I}_1} A_i \quad where \quad \mathcal{I}_1 = \{i:\varsigma(w_i^{(k)})=1\}\\
B_2 = \bigcup_{i \in \mathcal{I}_2} A_i \quad where \quad \mathcal{I}_2 = \{i:\varsigma(w_i^{(k)})=-1\}
\end{align} 
where $\varsigma(w_i^{(k)})=sign(w_i^{(k)})$. The sets $B_1$ and $B_2$ corresponding to the eigenvectors ${w}^{(k)}$, $k=2,3,4$ are presented in Figure \ref{bundle}.  We propose that regions characterised by the same colour (red and blue in our figures) are associated with separate bundles of UPOs. {\color{black}As we will see below, for each eigenvector, the red (blue) regions describe parts of the attractors with positive (negative) anomalies of the density with respect to the invariant one.} The forward trajectory undergoes transitions between the neighbourhood of the UPOs belonging to a bundle, and is repelled with low probability towards the  neighbourhood of an UPO belong to the other bundle. The closer to one the real part of an eigenvalue, the less efficient is the exchange between regions of different colours in the corresponding mode. More precisely, the subdominant eigenvectors ${w}^{(k)}$ provide an ordering of the quasi-invariant structures in terms of "leakiness". %In fact, we observe that finer quasi-invariant structures are associated to eigenvalues with smaller real part, which corresponds to faster mixing (Fig.\ref{fig: subfig 9b}, \ref{fig: subfig 9c}). 

{\color{black}Keeping in mind that each individual UPO is an actual invariant set and provides an exact \textcolor{black}{solution} of the evolution equations, we propose that our method defines structures that are closely related to the so-called quasi-invariant sets \cite{dellnitz_1999,dellnitz_1997,froyland_2005,froyland_2008,froyland_2003}. Loosely speaking, quasi-invariant sets are macroscopic dynamical \textcolor{black}{structures} such that the probability of individual trajectories beginning in the subset would leave it in short time is very little (see Appendix \ref{quasi} for more details). In particular, the red and blue regions in Figs. \ref{fig: subfig 9a}, \ref{fig: subfig 9b}, and \ref{fig: subfig 9c} closely resemble the structures defined by the first three Fiedler vectors defining the connectivity of the graph describing the mass transport of the L63 model (Figs. 5a,b and 6 in \cite{froyland_2001}).}

\begin{comment}

$B_1$ and $B_2$ - corresponding to the alternating red and blue regions -  identify two quasi-invariant structures in the phase space between which we expect little diffusion. %Specifically, we have that the alternating blue and red region correspond approximately to two separate quasi-invariant sets. \\
We now take a step further and extract information from the other subdominant eigenvalues $\lambda_k$ of the transfer operator. More precisely, the subdominant eigenvectors $\textbf{w}^{(k)}$ provide an ordering of the quasi-invariant structures in terms of "leakiness" (See e.g. \cite{froyland_2003,huisinga_2006,deuflhard_2005}) .
More precisely, the subdominant eigenvectors $\textbf{w}^{(k)}$ provide an ordering of the quasi-invariant structures in terms of "leakiness". In fact, we observe that finer quasi-invariant structures are associated to eigenvalues with smaller real part, which corresponds to faster diffusion (Fig.\ref{fig: subfig 9b}, \ref{fig: subfig 9c}) . This correspondence is interesting but not surprising, because the physical process responsible for the slow decay of anomalies of a probability measure with respect to the invariant one described in Fig. \ref{cubes} is indeed the slow mixing occurring in phase space between the regions described by the quasi-invariant sets depicted in Fig. \ref{bundle}.
\end{comment} 

\subsection{Relaxation Modes}

{\color{black}The red-and-blue representation of the subdominant modes given in Figs. \ref{fig: subfig 9a}-\ref{fig: subfig 9c} is essentially qualitative because we distinguish the various UPOs only in terms of the sign of their projection on the eigenvectors. We want now to portray the eigenmodes in $\mathbb{R}^3$, in such a way that it is possible to retain quantitative information associated to the evolution of ensembles of trajectories.} %order to understand the geometrical structure associated with the UPOs.
We  proceed as follows. {\color{black}
We partition %It is known \cite{sparrow_1982} that the attractor of the L63 model is contained in a 
the compact subset of $\mathbb{R}^3$ given by
the Cartesian product $\mathcal{P}=[-20,20]\times [-27.5,27,5]\times [1,48]$. As mentioned before, this set includes the attractor of the L63 model. We cover this region with $103400$ cubes  $\mathcal{D}=\{D_i\}_{i=1}^{103400}$ with sides having unitary length. The cubes are built having adjacent sides, so that $\mathcal{D}$ constitutes a partition of $\mathcal{P}$. Each UPO and corresponding neighbourhood intersects a certain number of cubes and each cube might contain contributions from different orbits. We now define a quantity (mass) that weights the contribution given by UPOs of different types within each cube. 
We set a fixed number of points $\bar{N}$ to be represented in the phase space a priori and assign the points to the different UPOs and relative neighbourhood  depending on the weight given by the corresponding component of the eigenvector $w^{(k)}$. These points are chosen along the orbits equally spaced in time. We also distinguish between negative and positive contributions, depending on the sign of the component $w^{(k)}_i$.}
{\color{black}We finally quantify the mass contained in each cube $D_i$ of the partition by calculating the algebric sum of the points contained in it.}

%We consider the transformation $\hat{w}^{(k)} = |w^{(k)}|/||w^{(k)}||_1$ and the mapping $\varsigma(w_i^{(k)})=sign(w_i^{(k)})$ provided by $\textbf{w}^{(k)}$. Each state $A_i$ will be awarded of a number of points $\bar{n}_i$ given by $\bar{n}_i = w_i^{(k)} \cdot \bar{N}$. We will sample these points uniformly along the UPO trajectory.  Thus, each states of the system (i.e. each periodic orbit and its neighbourhood $A_i$) will be characterised by two quantities: $\varsigma(w^{(k)}_i)$, its sign, and $\bar{n}_i$, its weight.
%Let then $\theta$ be a map $\theta: \mathcal{D}\to \mathbb{R}$  that associates a numerical value to each box of the partition, with the aim of quantifying the proportion of orbits of each kind. 

\begin{figure}
    \subfloat[]{%
      \includegraphics[width=7cm]{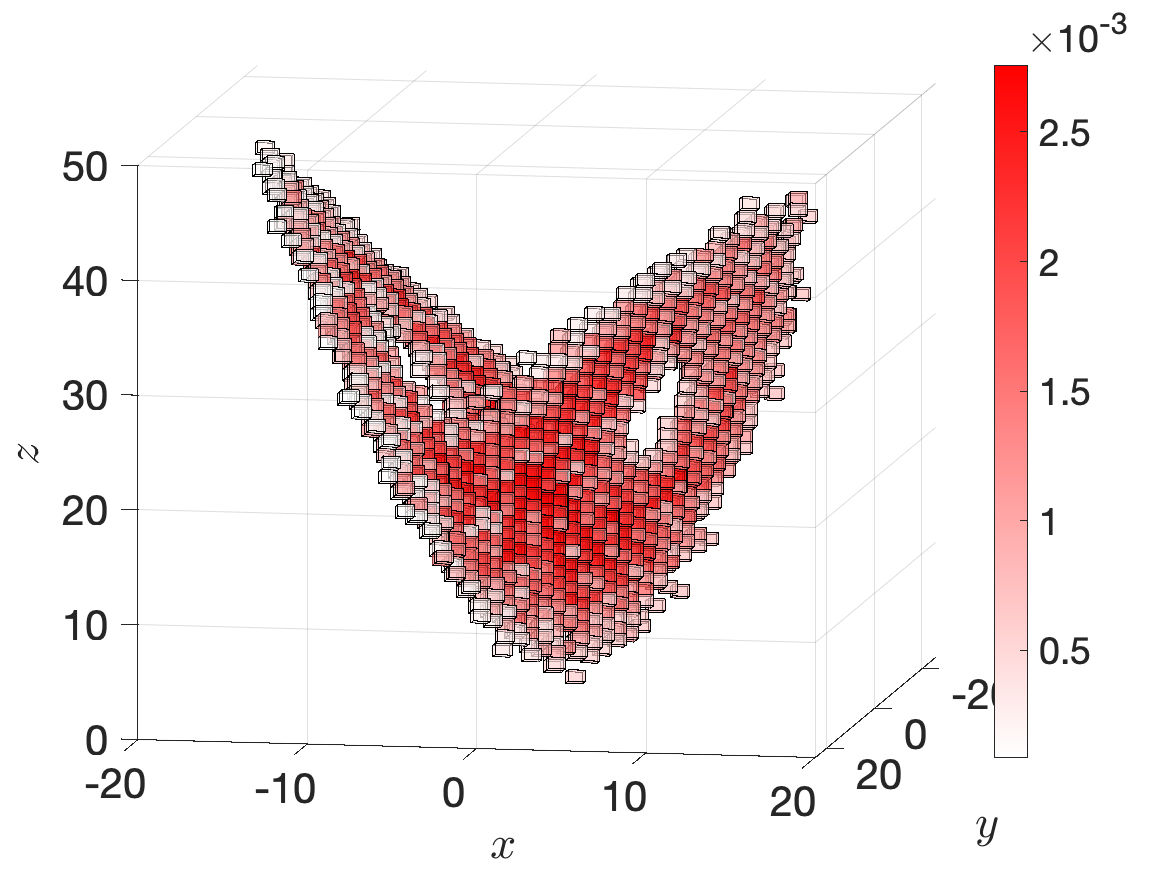}
           \label{fig: subfig 11a}
    }
    \qquad \qquad
     \subfloat[]{%
   \includegraphics[width=7cm]{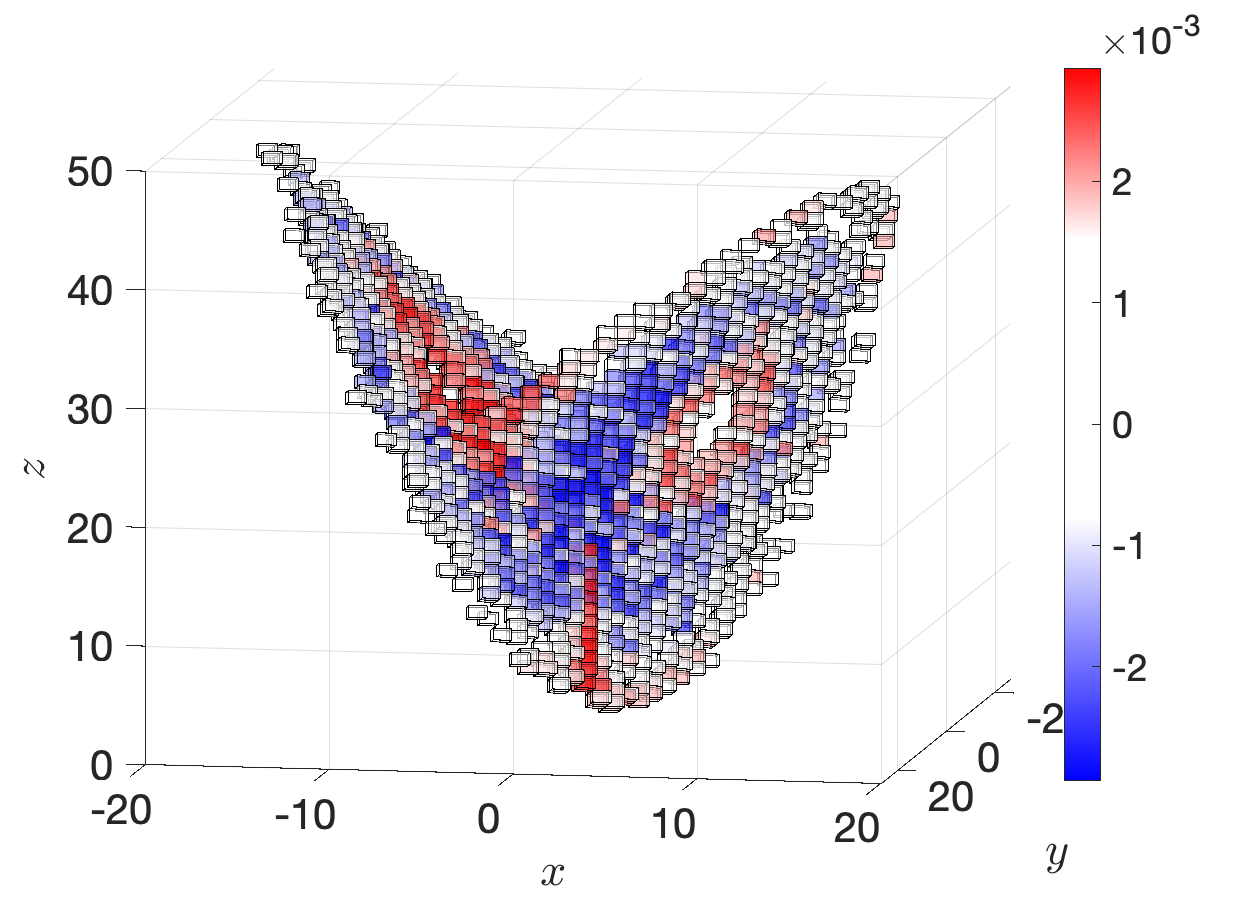}
              \label{fig: subfig 11b}
    }
    \qquad \qquad
         \subfloat[]{%
         \includegraphics[width=7cm]{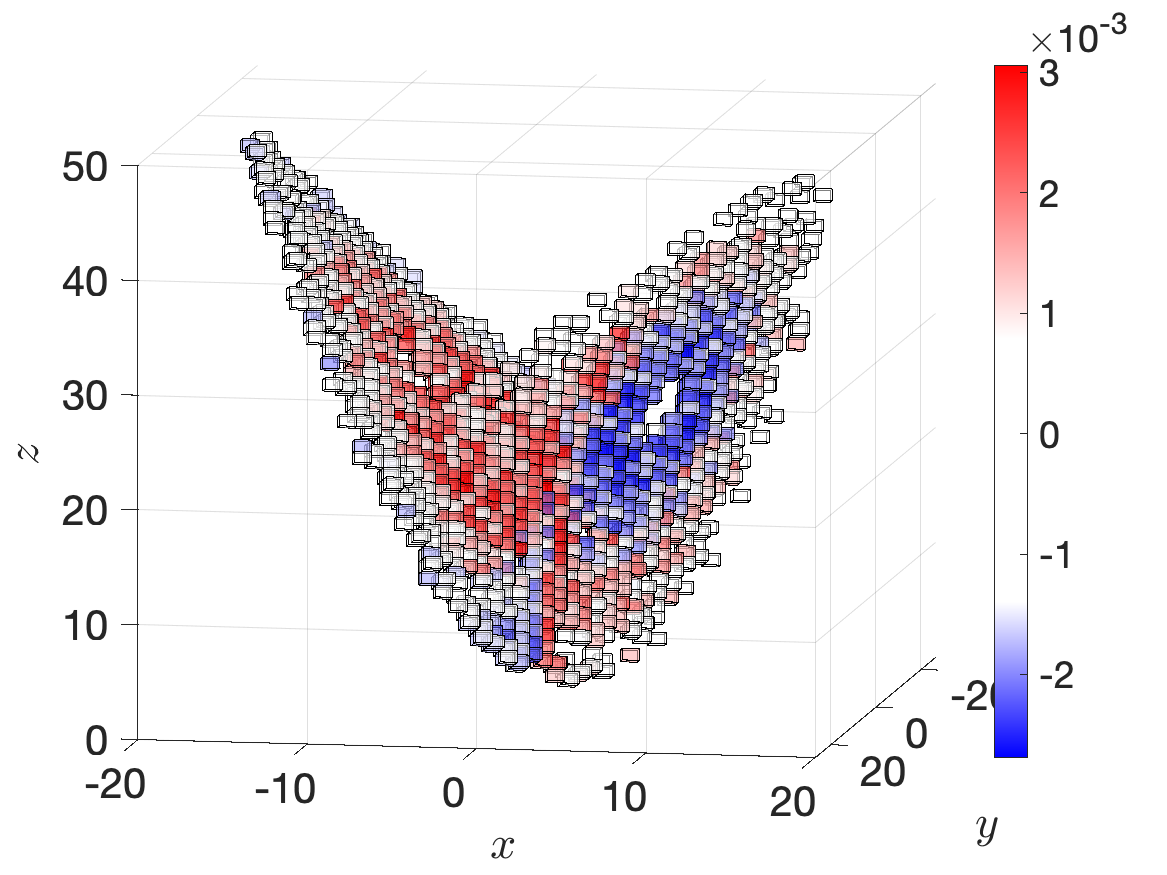}
            \label{fig: subfig 11c}
             }
    \qquad \qquad
         \subfloat[]{%
     \includegraphics[width=7cm]{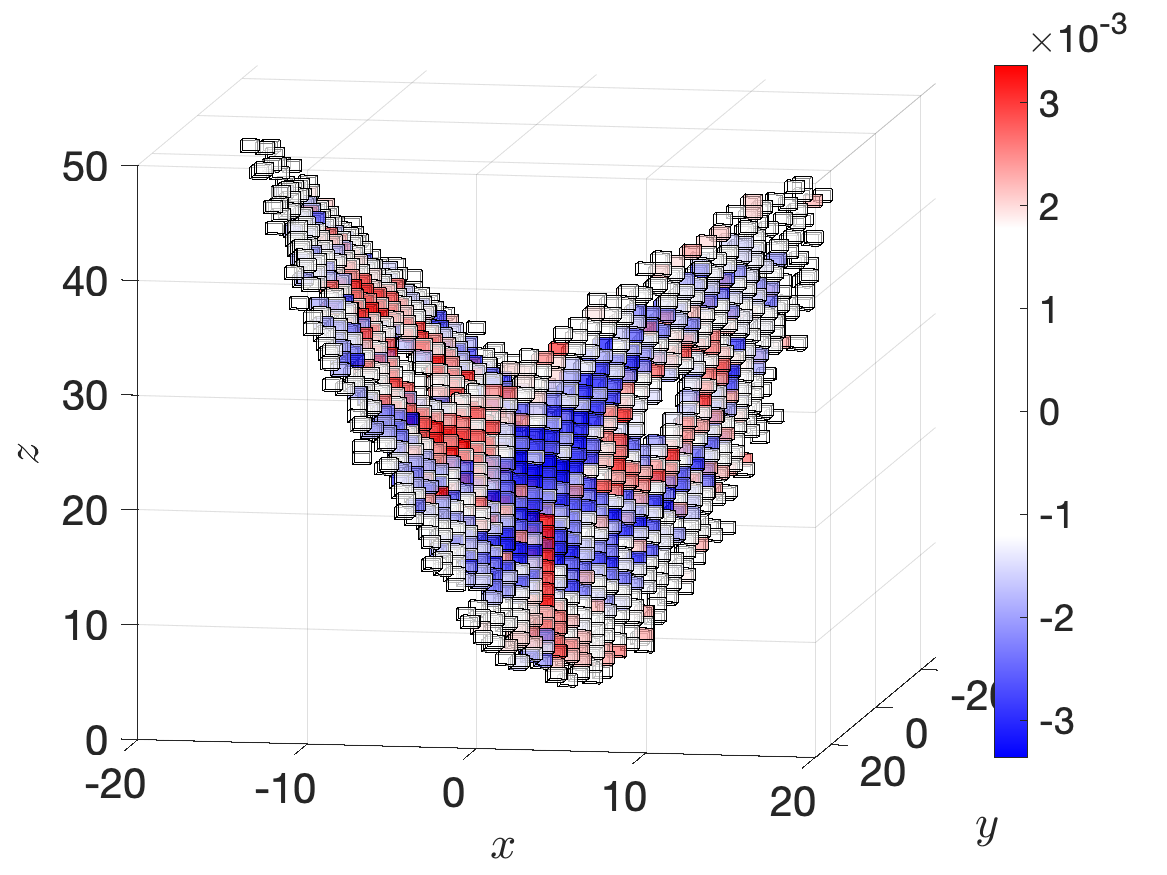}
                     \label{fig: subfig 11d}
                     }

    \caption{\label{cubes}Invariant Measure of the system obtained by projection of ${w}^{(1)}$ \protect\subref{fig: subfig 11a}. Projection in the phase space of \protect\subref{fig: subfig 11b}:  ${w}^{(2)}$ ($\lambda_2=0.9841$), \protect\subref{fig: subfig 11c}:  ${w}^{(3)}$ ($\lambda_3=0.9806$), \protect\subref{fig: subfig 11d}: ${w}^{(4)}$ ($\lambda_4=0.9706$).}
\end{figure}

%We consider the box $D_j$. Let us call $pos_j$ and $neg_j$ the number of positive and negative points contained in the box, obtained with the aforementioned procedure. We define $\theta(D_j)=pos_j+neg_j$, to quantify the mass contained in each cube. 
{\color{black}Correspondingly, Fig. \ref{fig: subfig 11a} describes the invariant measure, while Figs. \ref{fig: subfig 11b}, \ref{fig: subfig 11c}, and \ref{fig: subfig 11d} describe the eigenvectors corresponding to the subdominant eigenvalues $\lambda_2$, $\lambda_3$, and $\lambda_4$, respectively. The eigenvectors $w^{(2)}$, $w^{(3)}$, and $w^{(4)}$ are the three slowest modes responsible for the relaxation of an initial probability measure towards the invariant one, the rate of convergence being given by the corresponding eigenvalues. By construction, one can see a good correspondence between the red and blue regions in the panels of Figs. \ref{bundle} and \ref{cubes} associated with the same eigenvalue. 
%The forward trajectory undergoes transitions between the neighbourhood of the UPOs belonging to a bundle, and is repelled with low probability towards the  neighbourhood of an UPO belong to the other bundle. 
Indeed, the physical process responsible for the slow decay of anomalies of an ensemble  with respect to the invariant measure described in Fig. \ref{cubes} is indeed the slow mixing occurring in phase space between the regions described by the quasi-invariant sets associated with different bundles of UPOs depicted in Fig. \ref{bundle}. We observe that the smaller the eigenvalue, thus associated to faster decay rate, the finer the geometrical structure associated with the mode. This agrees with our intuition on how diffusion works. 
}

%We wish to provide a somewhat alternative interpretation of the plots corresponding to the subdominant eigenvalues as follows, keeping in mind that UPOs are invariant sets and provide exact solutions of the evolution equations. We propose that regions characterised by the same colour (red and blue in our figures) are associated with separate bundles of UPOs. The forward trajectory undergoes transitions between the neighbourhood of the UPOs belonging to a bundle, and is repelled with low probability towards the  neighbourhood of an UPO belong to the other bundle. The closer to one the real part of an eigenvalue, the less efficient is the exchange between regions of different colours in the corresponding mode.
%We observe that the smaller the eigenvalue, thus associated to faster decay rate, the finer the geometrical structure associated with the mode. This agrees with our intuition on how diffusion works.

\subsection{Remark}
{\color{black}The reader might wonder how robust the results presented in Figs. \ref{fig: subfig 9a}-\ref{fig: subfig 9c} and Figs. \ref{fig: subfig 11a}-\ref{fig: subfig 11d} with respect to the shadowing criteria defined in Eq. \ref{shadowing_definition}, which takes into consideration only tier 1 shadowing UPOs. To assess the robustness of the method, we have repeated our analysis using the looser definition of shadowing described in Sect. \ref{algorithm} that leads to increased persistence of the co-evolution of the chaotic trajectory and of the shadowing UPOs described in Fig. \ref{persistence}. The results are presented in the supplementary material. The subdominant eigenvectors change very little as larger values of $K$ are considered, whereas, as expected the value of the corresponding eigenvalues get closer and closer to 1, so that slower decay of correlation is found. Clearly, this is the probabilistic counterpart of the results shown in Fig. \ref{persistence} and supports the idea  expected, since allowing for more persistence in the shadowing of the chaotic trajectory results in less frequent transitions and thus slower decay rates.}

\section{Summary and Conclusion}
\label{conclusion}
The theory of UPOs has found extensive applications in the study of low-dimensional chaotic systems, in particular as a mean to calculate dynamical averages through the use of trace formulas \cite{eckhardt_1994,franceschini_1993,zoldi_1998}. In recent times promising developments {\color{black}have been made regarding its use for understanding the behaviour of higher dimension dynamical systems \cite{kawahara_2001,gritsun_2008,gritsun_2013,lucarini_2020,cvitanovic_2013,chandler_2013}. Very recently, efforts has been dedicated to better understanding the similarity of chaotic trajectory segments and of locally approximating UPOs in fluid flows \cite{yalniz_2020,krygier_2021}. It usually assumed that the low-period UPOs are the most relevant ones for achieving an accurate representation of statistical properties of the system \cite{eckhardt_1994,cvitanovic_1995,artuso_1990,Hunt1996,Yang2000}. Nonetheless, even if the trace formulas \cite{cvitanovic_1988} seem to suggest the opposite, it is sometimes found that long-period UPOs can be of great importance for computing statistical averages \cite{Zoldi1998,lasagna_2018,lasagna_2020}. Additionally, UPOs have been used as a way to perform coarse-graining: it has been shown that it is possible to approximate accurately the evolution of a fluid flow using a finite-state Markov chain where each state corresponds to the neighborhood of a UPOs \cite{Hof2021}. Finally, specific UPOs have been shown to key to separating quasi-invariant sets for the L63 model \cite{froyland_2009}. 

In this work we have attempted to bring together these research lines by performing an accurate analysis of how a long chaotic trajectory of the L63 model with the standard parameter values can be approximated using the complete set of UPOs having symbolic dynamics with period up to 14, numbering 2536 UPOs. The chaotic trajectory can be seen as a continuous process of scattering between the neighbourhood of the various UPOs. At each time step, we rank the UPOs in terms of their distance to reference point, and investigate how the distances and the  ranking changes in time. The shadowing of the trajectory involves both proximity and the fact that, as a result of the smoothness of the flow, the reference point of the trajectory and of the considered \textcolor{black}{UPOs co-evolve; indeed we can say that the rectified distance of the co-evolving UPO with the trajectory is of order of magnitude larger than the initial distance between the two. }
%The\textit{ ranked shadowing}, intended as the representation of approximation of a chaotic trajectory through the nearest UPOs, was at the center of our investigation which has been performed on L63. 
We find that longer UPOs, as a result of their higher number and longer spatial extent, are the most effective in shadowing the orbit of the system. This holds true if we consider a relaxed version of our  algorithm, which allow for the rank of the shadowing UPO to fluctuate up to a certain threshold (\textit{very good} vs. \textit{optimal} shadowing).

We then investigated a finite-state representation of the dynamics where each state is given by an UPO and its neighbourhood, and the stochastic matrix is defined in a frequentist way by studying the transitions defining the time-dependent shadowing of the chaotic trajectory. Since we are implementing a discretized representation of the transfer operator, the  eigenvectors corresponding to the subdominant eigenvalues describe the process of relaxation of ensembles towards the invariant measure. While a similar UPOs-based Markov chain model  has been recently proposed by \cite{Hof2021} with the goal of computing averages, to the best of our knowledge, this is the first time this specific discretization is performed  with the purpose of analysing the mixing properties of the system. By projecting the UPOs on the 3D space, we find that eigenvectors with finer spatial structures have faster decaying \textcolor{black}{rates}. Additionally, building on the fact that UPOs are  invariant sets that transport mass across the attractor, the regions of the eigenvectors having the same sign can be thought as approximately defining quasi-invariant sets. \textcolor{black}{Indeed, the patterns defined in this way exhibit qualitative agreement with the structures found in the L63 model by Froyland and Froyland and Padberg in \cite{froyland_2001} and \cite{froyland_2009} using the discretization of the transfer operator based on the classical Ulam's partition.} We interpret our findings as follows. The forward trajectory  typically undergoes scattering  between UPOs belonging to a bundle of UPOs associated with a quasi-invariant set, while, rarely, the scattering process bring the trajectory with close proximity of an UPO belonging to the other bundle, associated with a competing quasi-invariant set. \\

Clearly, further research is needed in this direction in order to assess differences and similarities between these approaches. Our procedure seems to have a good degree of robustness. It \textcolor{black}{is} encouraging to see that if we construct the stochastic matrix using the relaxed definition of the shadowing mentioned above, the eigenvectors corresponding to the subdominant eigenvalues are virtually unchanged, whereas the decay rates become slower, as persistence is enhanced by slowing down the transitions between the competing neighbourhoods. \\

%of the dynamics, defining the stochastic matrix controlling the shadowing of UPOs along the trajectory. In this context each UPO represents a possible state of the system. We found that this perspective on the dynamics allows for a new interpretation of quasi-invariant structures in terms of UPOs. By looking at the eigenvectors corresponding to subdominant eigenvalues we are able to explain on the one side the geometrical pattern associated with the decay of perturbations with respect to the invariant measure and on the other side link this behaviour to the structure of the invariant sets. 
This work provides further support to the potential of using UPOs for reaching a comprehensive understanding of the properties - averages and correlations - of chaotic dynamical systems. We would like to extend this analysis to higher dimensional system of practical relevance. In particular we would like to extend the work of Lucarini and Gritsun \cite{lucarini_2020} on blocking events, investigating transitions between zonal flow and blocking by applying the methodology developed in this paper. The investigation of this model is of interest both in terms of the  physical process of interest - the  low-frequency variability of the atmosphere is far from being a settled problem - and in terms of its mathematical properties, as it is characterised by high variability in the number of unstable dimension, thus featuring a serious violation of hyperbolicity.}

\appendix
%\label{app}
\section{Unstable Periodic Orbits Search}
\label{newton}
We will review here the classic Newton algorithm \cite{parker_2012} for detecting UPOs {\color{black}of the ordinary differential equation
\begin{equation}
\dot{x} = f(x),\quad x\in \mathcal{M}
\end{equation}
where $\mathcal{M}\subset\mathbb{R}^n$ is a compact manifold. This method is particularly appropriate for finding periodic solutions even in high-dimensional systems.} 

The problem of numerically finding UPOs can be reduced to the solution of the periodicity condition, which corresponds to a system of nonlinear equations with respect to the initial condition of the UPO and its period:
\begin{equation}
\label{periodicity_condition}
S^T(x_{in})=x_{in}.
\end{equation}where $x_{in}$ is the initial condition and $T$ is the period of the UPO.
Even for simple nonlinear systems this represents a difficult numerical problem. Hence, the choice of the algorithm and initial guess  represent an important aspect to be considered. We first rewrite the periodicity condition \ref{periodicity_condition} as follows:
\begin{equation}
S^T(x_{in}) -x_{in}=0
\end{equation}
 
This is a system of $n$ nonlinear equations (n is the dimension of the phase space) in $n+1$ unknowns (the vector $x_{in}$ and the orbit period T).
We start with an initial condition ($x_0$, T). A way to choose it is by calculating a long trajectory and selecting a quasi-recurrence occurring over a period T such that $|S^T(x_{in})-x_{in}|<\varepsilon$ with $\varepsilon$ decided a priori. 
Let then be $x^i$ and $T^i$ the $i$th approximations for initial condition and period. The aim of the algorithm is to calculate a correction $(\Delta x_i, \Delta T_i)$ so that we can improve the initial guess in such a way that
\begin{equation}
|| S^{T^i+\Delta T_i}(x^i + \Delta x_i)-(x^i + \Delta x_i)||<||  S^{T^i}(x^i)-x^i ||
\end{equation}
We obtain the approximate corrections $(\Delta x_i, \Delta T_i)$ by expanding 
\begin{align}
S^{T_{i+1}} (x^{i+1})- x_{i+1}  = S^{T_{i}+\Delta T_i} (x^{i}+\Delta x_i) -  (x^{i}+\Delta x_i)=0
\end{align}
into a Taylor series with respect to $\Delta x_i$ and $\Delta T_i$ 
\begin{widetext}
\begin{equation}\label{A5}
S^{T^{i}+\Delta T_i} (x^{i}+\Delta x_i)- (x^{i}+\Delta x_i) \approx S^{T^i}(x^i) - x^i+ \Big(\frac{\partial  S^{T^i}(y)}{\partial y}\Big|_{y=x^i}-I \Big)\Delta x_i + \frac{\partial  S^{T}(x^i)}{\partial T}\Big|_{T=T^i}\Delta T_i  = 0
\end{equation}
\end{widetext}
where $I$ is the identity matrix of order $n$. $\frac{\partial  S^{T_i}(y)}{\partial y}$ is the tangent linear operator and % and %fundamental matrix and it 
%represents the variation of $S^{T_i}(y)$ along $x_i$. 
it is an approximation $M_i$ of the monodromy matrix $M$ \cite{cvitanovic_2005}.  
$\frac{\partial  S^{T}(x_i)}{\partial T}\Big|_{T=T^i}$ is the derivative of the solution with respect to time $\dot x= f(x)$ evaluated at the final condition $f(S^{T_i}(x_i))$.
In order to remove the excess in degrees of freedom, we impose the \textit{phase condition} by requiring the orthogonality of the correction vector to the orbit
\begin{equation}
(f(S^{T_i}(x_i))) \cdot \Delta x_i = 0.
\end{equation}
In this way we reduce the problem of finding the corrections at step $i$ to the solution of a linear system of $n+1$ equations in $n+1$ unknowns
 \begin{equation}
 \begin{pmatrix}
M_i - I && f(S^{T_i}(x_i)) \\
(f(S^{T_i}(x_i)))^T  && 0
\end{pmatrix}
 \begin{pmatrix}
\Delta x_i \\
\Delta T_i
\end{pmatrix}
= 
 \begin{pmatrix}
 x_i - S^{T_i}(x_i) \\
0
\end{pmatrix}\label{A8}
\end{equation}
\begin{comment}
We define two errors associated to the numerical algorithm, them being at iteration $i$:
\begin{equation}
\label{errore_cond_in}
err^{in}_i := | S^{T_i}(x_i) - x_i|
\end{equation}
\begin{equation}
\label{error_correction}
err_i^{corr}:= | (\Delta x_i, \Delta T_i)| 
\end{equation}
We consider the UPO to be numerically detected when both \ref{errore_cond_in} and \ref{error_correction}  are sufficiently small.
\end{comment}
{\color{black}The solution of Eq. \ref{A8} gives the next approximations for the UPO initial condition and period. In some cases the Newton method may not give convergence (or the convergence could be very slow) if the initial guess is far from the solution, so that the linear Taylor expansion is not valid or the linear system is degenerate. In this case, one can use a nonlinear expansion in Eq. \ref{A5} as well as step relaxation together with line search procedure (see \cite{gritsun_2008} for more details). }

%\subsection{Choice of initial conditions}
We consider quasi-recurrent orbits as initial conditions. We integrate the system for a long time $T_{max}$ starting from a random initial state; the result is a numerical trajectory consisting of the set of ordered points $\{ x\}_{j=1}^{T_{max}}$. We then calculate the quantity $d_{ij} = |x_j-x_i|$ $\forall i,j \in \{1,...T_{max}\}$ and take the minimum, obtained at say $x_m, x_n$. We have a pair of points for which the trajectory starting from $x_m$ passes again near the starting point $x_m$ in time $n-m$. We can then consider the pairs $(x_m, m-n)$ as initial condition for determining the UPO  with the Newton method. 

%\subsection{Algorithm Specifications}
The numerical trajectories have been calculated using the midpoint numerical scheme, with integration time-step of $10^{-3}$. We choose an output time step $dt=0.01$ and consider a UPO to be detected when $err^{in}<\varepsilon$ with $\varepsilon = 10^{-10}$. 

\section{Quasi invariant sets}
\label{quasi}
We here introduce some key ideas regarding the macroscopic structures and large scale dynamics of the system. When the behaviour of individual trajectory is hard to predict, as it is the case in chaotic systems, the study of the global evolution of densities represents a powerful tool to gain insight into the dynamics. In fact, even if it is not possible to characterise the evolution of a single initial  condition, it often happens that we can group the phase space in sets characterised by predictable behaviour. 
Despite chaotic systems are often transitive, this property can be very weak and it is often the case that the phase space can be decomposed in macroscopic dynamical structure such that the probability of individual trajectories beginning in the subset would leave it in short time is very little. Trajectories tend to stay for a very long time in one of those regions before entering another region. We call these subsets quasi-invariant sets. More precisely, \cite{froyland_2009} let $\textbf{F}:\Omega\in\mathbb{R}^d:\to \mathbb{R}^d$ be a smooth vector field, generating the dynamical system or flow $\{\Phi^t\}_{t\in\mathbb{R}}$, $\Phi^t:\Omega \to \Omega $ be the flow of the autonomous system, $\mu$ preserved by $\Phi$. We say that a subset $A \subset \Omega$ is \textit{almost-invariant} over the interval $[0, \tau]$ if 
    
\begin{align}
\rho_{\mu, \tau} := \frac{\mu(A\cap \Phi_{-\tau}(A))}{\mu(A)}\approx 1
\end{align}
  
Quasi-invariant sets can also be regarded as a valuable tool to study transport and mixing properties of the flow \cite{froyland_2014}, by evolving with minimal dispersion.

\begin{acknowledgments}
The authors have benefitted from scientific exchanges with P. Cvitanovi\'c, J. Dorrington, G. Ducci, G. Froyland, C. Nesbitt, M. Santos, N. Zagli, M. Zaks and from the very constructive criticism by two anonymous reviewers. AG was supported by the Moscow Center of Fundamental and Applied Mathematics (Agreement 075-15-2019-1624 with the Ministry of Education and Science of the Russian Federation).VL acknowledges the support received from the EPSRC project EP/T018178/1 and from the EU Horizon 2020 project TiPES (grant no. 820970). CCM has been supported by an EPSRC studentship as part of the Centre for Doctoral Training in Mathematics of Planet Earth (grant number EP/L016613/1). {\color{black}The authors acknowledge the support received from  Institutional Sponsorship-International Partnerships-University of Reading EP/W524268/1.}
\end{acknowledgments}

\section*{Data Availability Statement}

The data that support the findings of this study, extra figures, videos, and further details on the methodology can be accessed through the project  "Decomposing the Dynamics of the Lorenz 1963 model using Unstable Periodic Orbits: Averages, Transitions, and Quasi-Invariant Sets" at \url{https://tinyurl.com/4z6hh9a3}.

\end{document}